\documentclass[twocolumn,prb,showpacs]{revtex4}
\usepackage{psfrag}
\usepackage{graphicx}
\usepackage{amsmath}
\usepackage{amssymb}
\usepackage{eufrak}
\usepackage{times}

\begin{document}
\title{Conserving $GW$ scheme for nonequilibrium quantum transport in
molecular contacts}

\author{Kristian S. Thygesen$^1$}\author{Angel Rubio$^2$}
\affiliation{$^1$Center for Atomic-scale Materials Design (CAMD), Department of Physics, Technical University of Denmark, DK - 2800 Kgs. Lyngby, Denmark.  \\
$^2$\mbox{European Theoretical Spectroscopy Facility (ETSF), Departamento de
F\'{\i}sica de Materiales, Edificio Korta,} \mbox{Universidad
del Pa\'{\i}s Vasco, Centro Mixto CSIC-UPV, and Donostia International
Physics Center (DIPC),} Avenida de Tolosa 71, E-20018 Donostia-San
Sebasti\'an, Spain.}
\date{\today}

\begin{abstract}
  We give a detailed presentation of our recent scheme to include
  correlation effects in molecular transport calculations using the
  non-equilibrium Keldysh formalism. The scheme is general and can be
  used with any quasiparticle self-energy, but for practical reasons
  we mainly specialize to the so-called $GW$ self-energy, widely used
  to describe the quasiparticle band structures and spectroscopic
  properties of extended and low-dimensional systems. We restrict the $GW$
  self-energy to the central region, and describe the leads by density functional theory (DFT). A minimal basis of
  maximally localized Wannier functions is applied both in the central $GW$
  region and the leads. The
  importance of using a conserving, i.e. fully self-consistent, $GW$ self-energy
  is demonstrated both analytically and by numerical examples. We
  introduce an effective spin-dependent interaction which
  automatically reduces self-interaction errors to all orders in the
  interaction. The scheme is applied to the Anderson model in- and out
  of equilibrium. In equilibrium at zero temperature we find that
  $GW$ describes the Kondo resonance fairly well for intermediate
  interaction strengths. Out of equilibrium we demonstrate that the
  one-shot $G_0W_0$ approximation can produce severe errors, in
  particular at high bias. Finally, we consider a benzene
  molecule between featureless leads. It is found that the molecule's
  HOMO-LUMO gap as calculated in $GW$ is significantly reduced as the
  coupling to the leads is increased, reflecting the more efficient
  screening in the strongly coupled junction. For the $IV$ characteristics of
  the junction we find that HF
  and $G_0W_0[G_{\text{HF}}]$ yield results closer to $GW$ than does DFT and $G_0W_0[G_{\text{DFT}}]$. This is explained in terms of self-interaction effects and
  life-time reduction due to electron-electron interactions.
\end{abstract}

\maketitle 

\section{Introduction} 
Since the first measurements of electron transport through
single molecules were reported in the late
nineties~\cite{agrait_report,reed97,smit02}, the theoretical interest
for quantum transport in nano-scale systems has been rapidly growing.
An important driving force behind the scientific developments is the
potential use of molecular devices in electronics and
sensor applications. On the other hand it is clear that a successful
introduction of these technologies is heavily dependent on the
availability of theoretical and numerical tools for the accurate
description of such molecular devices. 

So far, the combination of density functional theory (DFT) and
non-equilibrium Green's functions (NEGF) has been the most popular
method for modeling nano-scale conductivity~\cite{xue01,taylor01,brandbyge02,thygesen_bollinger03}. For strongly coupled
systems such as metallic point contacts, monatomic chains, and
contacts with small chemisorbed molecules, this combination has been
remarkably successful~\cite{sknielsen,djukic,strange}, but in the
opposite limit of weakly coupled systems where the conductance is much
smaller than the conductance quantum, $G_0=2e^2/h$, the NEGF-DFT method has
been found to overestimate the conductance relative to
experiments~\cite{stokbro03,heurich,bda}. Part of this discrepancy might
result from the use of inappropriate exchange-correlation (xc)
functionals~\cite{burke_evers}. However, it is important to remember that
the application of ground state DFT to non-equilibrium transport
cannot be rigorously justified - even with the exact xc-functional. In particular, a breakdown of the effective single-particle DFT description is expected
when correlation effects are important or when the system is driven
out of equilibrium.

Over the years several different schemes have
been proposed as alternatives to NEGF-DFT. Historically, the first
DFT based transport methods used an equivalent formulation in terms of
scattering states rather than Green's
functions~\cite{hirose94,lang95,choi99}. A more recent approach (still
within DFT), solves a master
equation for the density
matrix of an electron system exposed to a constant electric field and
coupled to a damping heat bath of auxiliary phonons~\cite{gebauer04}.

A few attempts have been made to calculate the current in the presence
of electronic correlations. In one approach the density matrix is
obtained from a many-body wave function and the non-equilibrium
boundary conditions are invoked by fixing the
occupation numbers of left- and right going
states~\cite{delaney}.
Exact diagonalization within the molecular subspace has been combined with rate
equations to calculate tunneling currents to first order in the
lead-molecule coupling strength~\cite{hettler}. The linear
response conductance of jellium quantum point contacts has been
addressed on the basis of the Kubo formula~\cite{bokes,malet}. Although this method is
restricted to the low bias regime, it has the advantage over the NEGF
method that interactions outside the device region can be naturally included. The time
dependent version of density functional theory has also been used as framework for quantum
transport~\cite{stefanucci_tddft,diventra_todorov,kurth}. This scheme is particularly
useful for simulating transients and high frequency
ac-responses. Within the NEGF formalism
the many-body $GW$ approximation has been used to address correlated transport both under
equilibrium~\cite{darancet} and non-equilibrium~\cite{thygesen_gw} conditions. 

Within the framework of many-body perturbation theory (MBPT)
electronic correlations are described by a self-energy which in
practice must be obtained according to some approximate scheme, e.g.
by summing a restricted set of Feynman diagrams. The important
question then arises whether the quantities calculated from the
resulting Green's function will obey the simple conservation laws. In
the context of quantum transport the continuity equation, which
ensures charge conservation, is obviously of special interest. An
elegant way of invoking the conservation laws is to write the
self-energy as the functional derivative of a so-called
$\Phi$-functional, i.e. $\Sigma[G]=\delta \Phi[G]/\delta G$. Since the
self-energy in this way becomes dependent on the Green's function
(GF), it must be determined self-consistently in conjunction with the
Dyson equation.~\cite{baym62}

Due to the large computational demands connected with the
self-consistent solution of the Dyson equation, practical $GW$ band
structure calculations usually evaluates the self-energy at some
approximate non-interacting $G_0$. This non self-consistent scheme
does not constitute a conserving approximation. While this might not
be important for the calculated spectrum, self-consistency has been
demonstrated to be fundamental for out-of-equilibrium
transport~\cite{thygesen_gw}. In addition to its conserving nature,
another nice feature of the self-consistent approach is that it leads
to a unique GF and thus removes the $G_0$-dependence inherent in the
non self-consistent approach.

A reliable description of electron transport through a molecular
junction requires first of all a reliable description of the
\emph{internal} electronic structure of the molecule itself, i.e. its
electron addition and removal energies. The $GW$ approximation has
been widely and successfully used to calculate such quasiparticle
excitations in both semi conductors, insulators and
molecules~\cite{rmp,hybertsen86,niehaus,louie,leeuwen06}, and on this basis it
seems natural to extend its use to transport calculations.

There are two main obstacles related to the extension of the $GW$
method to charge transport. First, the conventional application of the
$GW$ method has been to ground state problems whereas transport is an
inherent non-equilibrium problem. Secondly, it is not obvious how to
treat electron-electron interactions in the leads within the NEGF
formalism. In Ref.~\onlinecite{thygesen_gw} we proposed to overcome
these problems by extending the $GW$ self-energy to the Keldysh
contour and restricting it to a finite central region where
correlation effects are expected to be most important. In the
present paper we provide an extended presentation of these ideas.

When a molecule is brought into contact with electrodes a number of
physical mechanisms will affect its electronic structure.  Some of
these mechanisms are single-particle in nature and are already well
described at the DFT Kohn-Sham level. But there are also important
many-body effects which require a dynamical treatment of the
electronic interactions. One example is the renormalization of the
HOMO-LUMO gap induced by the image charges formed in the electrodes
when an electron is added to or removed from the
molecule.~\cite{louie,kubatkin} Another example is the Kondo effect
which results from correlations between a localized spin on the
molecule and delocalized electrons in the
electrodes~\cite{goldhaber,costi94}. Third, as we will show here, the
coupling to (non-interacting) electrodes enhances the screening on the
molecule leading to acharacteristic reduction of the HOMO-LUMO gap as
function of the electrode-molecule coupling strength.

In this paper, we focus on improving the description of quantum
transport in molecular junctions by improving the description of the internal
electronic structure of the molecule while preserving a
non-perturbative treatment of the coupling to leads. We do this within the NEGF
formalism by using a self-consistent $GW$
self-energy to include xc effects within the molecular
subspace which in turn is coupled to non-interacting leads. The rationale behind this division 
is that the transport properties to a large extent are determined by the
narrowest part of the conductor, i.e. the molecule, while the leads
mainly serve as particle
reservoirs. Strictly speaking this is correct only when a 
sufficiently large part of the leads is included in the
$GW$ region. If the central region is too small, spurious
back-scattering at the interface between the $GW$ and
mean-field regions might affect the calculated conductance. Furthermore, the dynamical
formation of image charges in the electrodes requires that part of the
electrodes are included in the $GW$ region. In the present work we do,
however, not attempt to address this latter effect.

The paper is organized as follows. In Sec.~\ref{sec.formalism} we
introduce the model used to describe the transport problem and review
the basic elements of the Keldysh Green's function formalism. In
Sec.~\ref{sec.gw} we introduce an effective interaction, discuss the
problem of self-interaction correction in diagrammatic expansions, and
derive the non-equilibrium $GW$ equations for an interacting region
coupled to non-interacting leads. In Sec.~\ref{sec.conserving} we
introduce the current formula and show that charge conservation is
fulfilled within the NEGF formalism for $\Phi$ derivable self-energies
- also when incomplete basis sets are used. The practical
implementation of the $GW$ transport scheme using a Wannier function
basis obtained from DFT is described in Sec. \ref{sec.method}. In
Secs.~\ref{sec.anderson} and \ref{sec.benzene} we present results for
the non-equilibrium transport properties of the Anderson impurity
model and benzene molecule between jellium leads, respectively. We
conclude in Sec.~\ref{sec.conclusions}

\section{General Formalism}\label{sec.formalism}
  In this section we review the elements of the Keldysh Green's
  function formalism necessary to deal with the non-equilibrium
  transport problem. To limit the technical details we
  specialize to the case of orthogonal basis sets and
  refer to Ref.~\onlinecite{nonorthogonal} for a 
  generalization to the non-orthogonal case. 

\subsection{Model}\label{sec.model}
We consider a quantum conductor consisting of a central region ($C$)
connected to left ($L$) and right ($R$) leads. For times
  $t<t_0$ the three regions are decoupled from each other, each being
  in thermal equilibrium with a common temperature, $T$, and chemical
  potentials $\mu_L,\mu_C$, and $\mu_R$, respectively. At $t=t_0$ the coupling between the three subsystems is
  switched on and a current starts to flow as the electrode with higher
  chemical potential discharges through the central region into the
  lead with lower chemical potential. Our aim is to calculate the
  steady state current which arise after the transient has died out.

We denote by $\{\phi_i\}$ an orthonormal set of
single-particle orbitals, and by $\mathcal{H}$ the Hilbert space
spanned by $\{\phi_i\}$. The orbitals $\phi_i$ are assumed to be
localized such that $\mathcal{H}$ can be decomposed into a sum of
  orthogonal subspaces corresponding to the division of the system
  into leads and central region, i.e. $\mathcal H = \mathcal H_L+\mathcal H_C+\mathcal H_R$. We will
  use the notation $i\in \alpha$ to indicate that $\phi_i\in
  \mathcal{H}_{\alpha}$ for some $\alpha \in \{L,C,R\}$. 

The non-interacting part of the Hamiltonian of the \emph{connected} system is written
\begin{equation} \label{eq.nonintham}
\hat h=\sum_{{i,j\in}\atop {L,C,R}}\sum_{\sigma=\uparrow \downarrow}h_{ij}c_{i\sigma}^{\dagger}c_{j\sigma}
\end{equation}
where $i,j$ run over all basis states of the system. For
$\alpha,\beta\in\{L,C,R\}$, the operator $\hat h_{\alpha \beta}$ is
obtained by restricting $i$ to region $\alpha$ and $j$ to region
$\beta$ in Eq.~(\ref{eq.nonintham}). Occasionally we shall write $\hat
h_\alpha$ instead of $\hat h_{\alpha \alpha}$. We assume that there is no direct
coupling between the two leads, i.e. $\hat h_{LR}=\hat h_{RL}=0$ (this
condition can always be fulfilled by increasing the size of the
central region since the basis functions are localized). We
introduce a special notation for the "diagonal" of $\hat h$,
\begin{equation}
\hat h_0=\hat h_{LL}+\hat h_{CC}+\hat h_{RR}.
\end{equation}
It is instructive to note that $\hat h_0$ does \emph{not} describe the three
regions in isolation from each other, but rather the contacted
system without inter-region hopping.
We allow for interactions between electrons inside the central
region. The most general form of such a two-body interaction is, 
\begin{equation}\label{eq.interaction}
\hat V=\sum_{{ijkl\in C} \atop {\sigma\sigma'}} V_{ij,kl}c^{\dagger}_{i\sigma}c^{\dagger}_{j\sigma'}c_{l\sigma'} c_{k\sigma}.
\end{equation}
The full Hamiltonian describing the system at time $t$ can then be written
\begin{eqnarray}\label{eq.fullH}
\hat H(t)=\left \{ \begin{array}{ll}
\hat H_0= \hat h_0+\hat V& \text{for }t<t_0\\ 
\hat H = \hat h+\hat V& \text{for }t>t_0
\end{array} \right.
\end{eqnarray}
Notice, that we use small letters for non-interacting quantities while
the subscript 0 refers to uncoupled quantities.
The specific form of the matrix elements $h_{ij}$ and $V_{ij,kl}$
defining the Hamiltonian are considered in Sec.~\ref{sec.method}.

Having defined the Hamiltonian we now consider the intial state of the
system, i.e. the state at times $t<t_0$. For such times the three
subsystems are each in thermal equilibrium and thus characterized by
their equilibrium density matrices. For the left lead we have
\begin{equation}
\hat \varrho_L=\frac{1}{Z_L}\exp(-\beta(\hat h_{L}-\mu_{L}\hat N_{L}))
\end{equation}
with
\begin{equation}
Z_L=\text{Tr}[\exp(-\beta(\hat h_{L}-\mu_{L}\hat N_{L}))].
\end{equation}
Here $\beta$ is the inverse temperature and $\hat
  N_L=\sum_{\sigma,i\in L}c^{\dagger}_{i\sigma}c_{i\sigma}$ is the
  number operator of lead $L$.  $\hat \varrho_R$ and $Z_R$ are
  obtained by replacing $L$ by $R$. For $\hat \varrho_C$ and $Z_C$ we
  must add $\hat V$ to account for correlations in the initial state
  of the central region. The initial state of the whole system is then
  given by 
\begin{equation}\label{eq.initialstate}
\hat \varrho=\hat \varrho_{L}\hat \varrho_{C}\hat \varrho_{R},
\end{equation}
If $\hat V$ is not included in $\hat \varrho_C$ we obtain the
uncorrelated (non-interacting) initial state $\hat \varrho_{ni}$. We
note that the order of the density matrices in
Eq.~(\ref{eq.initialstate}) plays no role since they all commute due
to the orthogonality of the system $\{\phi_i\}$.  Because $\hat H_0$
($\hat h_0$) describes the contacted system without inter-region
hopping, $\hat \varrho$ ($\hat \varrho_{ni}$) does not describe the
three regions in physical isolation. In other words the three regions
are only decoupled at the \emph{dynamic} level for times
$t<t_0$.

\begin{figure}[!h]
  \includegraphics[width=0.88\linewidth]{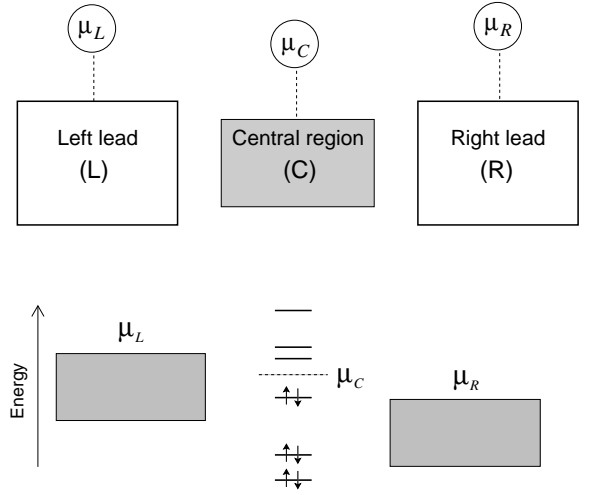}
\caption[cap.wavefct]{\label{fig_before} Before the coupling between the
  three regions is established, the three subsystmes are in
  equilibrium with chemical potentials $\mu_L$, $\mu_C$, and $\mu_R$,
  respectively. }
\end{figure}

\subsection{The Contour-ordered Green's function}
In this section we introduce the contour-ordered GF which is the
central object for the many-body perturbation theory in
non-equilibrium systems. For more detailed accounts of the NEGF theory
we refer to Refs.~\onlinecite{haug_jauho,leeuwen_dahlen}.
  
The contour-ordered GF relevant for the model
introduced in the previous section, is defined by
\begin{equation}\label{eq.gfdef}
G_{i\sigma,j\sigma'}(\tau,\tau')=-i\text{Tr}\{\hat \varrho T[c_{H,i\sigma}(\tau)c^{\dagger}_{H,j\sigma'}(\tau')]\}.
\end{equation}
Here $\tau$ and $\tau'$ are points on the Keldysh contour, $\mathcal
C$, which runs along the real time axis from $t_0$ to $\infty$ and
back to $t_0$, and $T$ is
the time-ordering operator on the contour. The creation and
annihilation operators are taken in the Heisenberg picture with
respect to the full Hamiltonian in Eq. (\ref{eq.fullH}). We do not
consider spin-flip processes and thus suppress the spin indices in the
following. 

In order to obtain an expansion of $G_{ij}(\tau,\tau')$ in powers of
$\hat V$, we switch to the interaction picture where we have
\begin{equation}\label{eq.gijint}
G_{ij}(\tau,\tau')=-i\text{Tr}\{\hat \varrho
  T[e^{-i\int_{\mathcal C}\text{d}\bar \tau \hat V_{h}(\bar \tau)}
  c_{h,i}(\tau) c^{\dagger}_{h,j}(\tau')]\}.
\end{equation}
By extending $\mathcal C$ into the complex plane by a
vertical branch running from $t_0$ to $t_0-i\beta$, we can replace
$\hat \varrho$ by the uncorrelated $\hat
\varrho_{ni}$~\cite{haug_jauho}. Neglecting the vertical branch then
corresponds to neglecting correlations in the central region's initial
state. While it must be expected that the
presence of initial correlations will influence the transient behavior
of the current, it seems plausible that they will be washed out over
time such that the steady state current will not depend on $\hat
\varrho_C$. Furthermore, in the
special case of equilibrium ($\mu_L=\mu_C=\mu_R$) and zero temperature, the
Gellman-Low theorem ensures that the correlations are
correctly introduced when starting from the uncorrelated initial state
at $t_0=-\infty$.\cite{fetterwalecka}
In practice the
neglect of initial correlations is a major simplification which allows
us to work entirely on the real axis avoiding any reference to the
imaginary time. For these reasons we shall adopt this approximation
and neglect initial correlations in the rest of this paper. 

Eq.~(\ref{eq.gijint}) with $\hat \varrho$ replaced by $\hat \varrho_{ni}$
constitute the starting point for a systematic series expansion of $G_{ij}$ in
powers of $\hat V$ and the free propagator, 
\begin{equation}
g_{ij}(\tau,\tau')=-i\text{Tr}\{\hat \varrho_{ni}T[ c_{h,i}(\tau) c^{\dagger}_{h,j}(\tau')]\},
\end{equation}
which describes the non-interacting electrons in the coupled system. 
The diagrammatic expansion leads to the identification of a
self-energy, $\Sigma$, which relates the interacting GF to the
non-interacting one through Dyson's equation 
\begin{equation}\label{eq.dyson1}
G(\tau,\tau')=g(\tau,\tau')+\int_{\mathcal C} \text{d} \tau_1 \text{d} \tau_2  g(\tau,\tau_1)\Sigma(\tau_1,\tau_2)G(\tau_2,\tau'),
\end{equation}
(matrix multiplication is implied). As we will see in Sec.~\ref{sec.current}, only the Green's function of the central region is needed for
the calculation of the current, and we can therefore focus on the
central-region submatrix of $G$. Due to the structure of $\hat V$, the
self-energy matrix, $\Sigma_{ij}$, will be non-zero only when
both $i,j\in C$, and for this reason $C$ subscripts can be added to all 
matrices in Eq.~(\ref{eq.dyson1}). Having observed this we will
nevertheless write $\Sigma$ instead of $\Sigma_C$ for notational simplicity.

The free propagator, $g_C(\tau,\tau')$, which is still a
non-equilibrium GF, satisfies the following Dyson equation
\begin{eqnarray}\label{eq.dyson3}\nonumber
g_C(\tau,\tau') =g_{0,C}(\tau,\tau')+\int_{\mathcal C} \text{d} \tau_1 \text{d} \tau_2  g_{0,C}(\tau,\tau_1) \\
\lbrack \Sigma_L(\tau_1,\tau_2)+\Sigma_R(\tau_1,\tau_2)\rbrack g_C(\tau_2,\tau'),
\end{eqnarray}
where $g_0$ is the \emph{equilibrium} GF defined by $\hat \varrho_{ni}$ and $\hat h_0$. The coupling self-energy due to lead $\alpha=L,R$ is given by
\begin{equation}\label{eq.sigmacontour}
\Sigma_{\alpha}(\tau,\tau')=h_{C\alpha}g_{0,\alpha}(\tau,\tau')h_{\alpha C}.
\end{equation}
Notice the slight abuse of notation: $\Sigma_{\alpha}$ is \emph{not}
the $\alpha\alpha$ submatrix of $\Sigma$. In fact $\Sigma_L$ and
$\Sigma_R$ are both matrices in the central region indices. Combining Eqs.~(\ref{eq.dyson1}) and (\ref{eq.dyson3}) we can write 
\begin{eqnarray}\label{eq.dyson4}\nonumber
G_C(\tau,\tau')&=&g_{0,C}(\tau,\tau')\\&+&\int_{\mathcal C} \text{d} \tau_1 \text{d} \tau_2  g_{0,C}(\tau,\tau_1)\Sigma_{tot}(\tau_1,\tau_2)G_C(\tau_2,\tau'),
\end{eqnarray}
which expresses $G_C$ in terms of the equilibrium propagator of the non-interacting,
uncoupled system, $g_0$, and the total self-energy
\begin{equation}
\Sigma_{tot}=\Sigma+\Sigma_L+\Sigma_R.
\end{equation}

\subsection{Real-time Green's functions}
  In order to evaluate expectation values of single-particle
  observables we need the real-time correlation functions. We work
  with two correlation functions also called the lesser and greater
  GFs and defined as 
\begin{eqnarray}\label{eq.gl}
G_{ij}^<(t,t')&=&i\text{Tr}\{\hat \varrho_{ni}
c^{\dagger}_{H,j}(t')c_{H,i}(t)\}\\ \label{eq.gg}
G_{ij}^>(t,t')&=&-i\text{Tr}\{\hat \varrho_{ni} c_{H,i}(t)c^{\dagger}_{H,j}(t')\}.
\end{eqnarray} 
Two other important real-time GFs
are the retarded and advanced GFs defined by
\begin{eqnarray}\label{eq.ret}
G_{ij}^r(t,t')&=&\theta(t-t')(G_{ij}^>(t,t')-G^<_{ij}(t,t'))\\ \label{eq.adv}
G_{ij}^a(t,t')&=&\theta(t'-t)(G_{ij}^<(t,t')-G^>_{ij}(t,t')).
\end{eqnarray} 
The four GFs are related via
\begin{equation}\label{eq.fundrelation}
G^>-G^<=G^r-G^a.
\end{equation}
The lesser
and greater GFs are just special cases of the contour-ordered GF. For
example $G^<(t,t')=G(\tau,\tau')$ when $\tau=t$ is on the upper branch
of $\mathcal C$ and $\tau'=t'$ is on the lower branch. This can be used
to derive a set of rules, sometimes referred to as the Langreth rules,
for converting expressions involving contour-ordered quantities into
equivalent expressions involving real-time quantities. We shall not
list the conversion rules here, but refer to
Ref.~\onlinecite{haug_jauho} (no initial correlations) or
Ref.~\onlinecite{leeuwen_dahlen} (including initial correlations). The usual
procedure in non-equilibrium is then to derive the relevant equations
on the contour using the standard diagrammatic techniques, and
subsequently converting these equations to real time by means of the
Langreth rules. An example of this procedure is given in
Sec.~\ref{sec.noneqgw} where the non-equilibrium $GW$
equations are derived.

\subsubsection{Equilibrium}
In equilibrium, the real-time GFs depend only on the time difference
$t'-t$. Fourier transforming with respect to this time difference then brings out the spectral
properties of the system. In particular the spectral
function 
\begin{equation}\label{eq.spectral}
A(\omega)=i[G^r(\omega)-G^a(\omega)]=i[G^>(\omega)-G^<(\omega)]
\end{equation} 
shows peaks at the quasiparticle (QP) energies of the system. In
equilibrium we furthermore have the fluctuation-dissipation
theorem,
\begin{eqnarray}\label{eq.fluct1}
G^<(\omega)&=&if(\omega-\mu)A(\omega)\\
G^>(\omega)&=&-i(1-f(\omega-\mu))A(\omega), 
\end{eqnarray}
relating the correlation
functions to the spectral function and the Fermi-Dirac distribution function,
$f$. The fluctuation-dissipation theorem follows from the Lehman
representation which no longer holds out of equilibrium, and as a
consequence one has to work explicitly with the correlation functions
in non-equilibrium situations.

\subsubsection{Non-equilibrium steady state}
We shall work under the assumption that in steady state, all the
real-time GFs depend only on the time-difference $t'-t$. Taking the limit $t_0\to -\infty$
this will allow us to use the
Fourier transform to turn convolutions in real time into products in
frequency space. Applying the Langreth conversion rules to the Dyson
equation (\ref{eq.dyson4}), and Fourier transforming with respect to
$t'-t$ then leads to the following expression for the retarded GF of the central region
\begin{equation}
G^r_C(\omega)= g_{0,C}^r(\omega)+g_{0,C}^r(\omega)\Sigma^r_{tot}(\omega)G^r_C(\omega).
\end{equation}
This equation can be inverted to yield the closed form
\begin{equation}\label{eq.grc_closed}
G^r_C(\omega)= [(\omega+i\eta)I_C-h_{C}-\Sigma^r_L(\omega)-\Sigma^r_R(\omega)-\Sigma^r(\omega)]^{-1}.
\end{equation}
The equation for $G^a$ is obtained by replacing $r$ by $a$ and $\eta$
by $-\eta$ or, alternatively, from $G^a=(G^r)^{\dagger}$.
For the lesser correlation function the conversion rules lead to the expression
\begin{equation}\label{eq.keldysh}
G_C^{</>}= G_C^r\Sigma^{</>}_{tot}G_C^a(\omega)+\Delta^{</>}
\end{equation}
where
\begin{equation}\label{eq.delta}
\Delta^{</>}=[I_C+G_C^r\Sigma^r_{tot}]g_{0,C}^{</>}[I_C+\Sigma^a_{tot}G_C^a].
\end{equation}
The $\omega$-dependence has been suppressed for notational simplicity. Using that
$\Sigma_{tot}^{r/a}=(g_{0,C}^{r/a})^{-1}-(G_C^{r/a})^{-1}$ together with the equilibrium
relations $g_{0,C}^<=-f(\omega-\mu_C)[g_{0,C}^r-g_{0,C}^a]$ and $g_{0,C}^>=-(f(\omega-\mu_C)-1)[g_{0,C}^r-g_{0,C}^a]$, we
find that
\begin{eqnarray}\label{eq.delta1}
\Delta^<(\omega)&=&2i\eta f(\omega-\mu_C)G_C^r(\omega)G_C^a(\omega)\\ \label{eq.delta2}
\Delta^>(\omega)&=&2i\eta [f(\omega-\mu_C)-1]G_C^r(\omega)G_C^a(\omega).
\end{eqnarray}
If the product $G^r(\omega) G^a(\omega)$ is independent of $\eta$ we can conclude that $\Delta(\omega) \to 0$
in the relevant limit of small $\eta$. However, as explained below, this is not always the case. 

\subsubsection{Bound states and the $\Delta$-term}\label{sec.boundstates}
We first focus on non-interacting electrons. In this case the
non-equilibrium correlation functions
$g^{</>}$ must be evaluated from Eq.~(\ref{eq.keldysh}) with
$\Sigma_{tot}=\Sigma_L+\Sigma_R$. For energies outside the band-width
of the leads we have $\Sigma_{\alpha}^r-\Sigma_{\alpha}^a=0$ such that
no broadening of the (non-interacting) levels is introduced by the
coupling to the leads. At such energies we have $g_C^r-g_C^a=2i\eta
g_C^r g_C^a$, and we conclude from
Eqs. (\ref{eq.delta1}),(\ref{eq.delta2}) that $\Delta^{</>}$ becomes
proportional to the spectral function, $A=g_C^r-g_C^a$. Since
$A(\omega)$ does not necessarily vanish outside the band-width of the leads
(it has delta peaks at the position of bound states), it follows
that $\Delta^{</>}$ should be included in the calculation of $g^{</>}$
to properly account for the bound states. It is interesting to notice
that $\mu_C$, which defines the initial state of the central region, drops
out of the equations for $g$ if and only if there are no bound states. 

When interactions are present in the central region correlation
effects will reduce the lifetime of any single-particle state in
$C$. Mathematically, this is expressed by 
the fact that $\Sigma^r-\Sigma^a$ will be non-zero
for all physically relevant energies. Consequently, the
product $G^r(\omega) G^a(\omega)$ will approach a finite value as $\eta \to 0$ leading to a vanishing $\Delta^{</>}$.

In conclusion, the $\Delta$ terms of
Eqs.~(\ref{eq.delta1}),(\ref{eq.delta2}) always vanish when
interactions are present in $C$, while for non-interacting electrons they
vanish everywhere except for $\omega$ corresponding to bound states.
We mention that it has recently been shown in the time-dependent NEGF
framework that the presence of bound states can affect the long time
behavior of the current in the non-interacting case~\cite{gianluca_bound}.

\section{The GW equations}\label{sec.gw}
In this section we derive and discuss the non-equilibrium $GW$ and
second order Born (2B) approximations. However, before addressing the
expressions for the self-energies
we introduce an effective interaction which leads to a particularly
simple form of the equations and at the same time provides a
means for reducing self-interaction errors in higher order
diagrammatic expansions.
 
\subsection{Effective interaction}\label{sec.effint}
The direct use of the full interaction Eq.~(\ref{eq.interaction})
results in a four-index polarization a function. The numerical representation
and storage of
this frequency dependent four-index function is very demanding,
and for this reason we consider the effective interaction defined by
\begin{equation}\label{eq.intapprox}
\hat V_{\text{eff}} = \sum_{ij,\sigma\sigma'} \tilde V_{i\sigma,j\sigma'}c^{\dagger}_{i\sigma}c^{\dagger}_{j\sigma'}c_{j\sigma'} c_{i\sigma},
\end{equation}
where 
\begin{equation}\label{eq.effint}
\tilde V_{i\sigma,j\sigma'}=V_{ij,ij}-\delta_{\sigma\sigma'}V_{ij,ji}.
\end{equation}
This expression follows by restricting the sum in the full interaction
Eq.~(\ref{eq.interaction}) to terms of the form
$V_{ij,ij}c^{\dagger}_{i\sigma}c^{\dagger}_{j\sigma'}c_{j\sigma'}
c_{i\sigma}$ and $V_{ij,ji}c^{\dagger}_{i\sigma}c^{\dagger}_{j\sigma}c_{j\sigma}
c_{i\sigma}$.

The effective interaction is local in orbital space, i.e. it is a two-point function instead
of a four-point function and thus resembles the real-space
representation. Note, however, that in contrast to the real-space
representation $\tilde
V_{i\sigma,j\sigma'}$ is spin-dependent. In particular the
self-interactions, $\tilde
V_{i\sigma,i\sigma}$, are zero by construction and consequently
self-interaction (in the orbital basis) is avoided to all orders in a perturbation expansion in powers of $\tilde
V$. Since the off-diagonal elements ($i\neq j$) of the exchange
integrals $V_{ij,ji}$ are small, on expects that the main effect of the second
term in Eq.~(\ref{eq.effint}) is to cancel the self-interaction in the
first term. 

It is not straightforward to anticipate the quality of a $GW$
calculation based on the
effective interaction (\ref{eq.intapprox}) as compared to the full
interaction (\ref{eq.interaction}). Clearly, if we include all
Feynman diagrams in $\Sigma$, we obtain the exact result when the
full interaction (\ref{eq.interaction}) is used, while the use of the effective interaction  
(\ref{eq.intapprox}) would yield an approximate result. The quality of
this approximate result would then depend on the basis
set, becoming better the more localized the basis functions and equal
to the exact result in the limit of completely localized delta
functions where only the direct Coulomb integrals $V_{ij,ij}$ will be
non-zero. 

However,
when only a subset of all diagrams are included in $\Sigma$ the
situation is different: In the $GW$ approximation only one
diagram per order (in $\hat V$) is included, and thus cancellation of
self-interaction does not occur when the full interaction is used. On
the other hand the effective interaction
(\ref{eq.effint}) is self-interaction free (in the orbital basis) by
construction. The situation can be understood by considering the
lowest order case. There are only two first order diagrams - the
Hartree and exchange diagrams - and each cancel the self-interaction
in the other. More generally, the presence of self-interaction in an
incomplete perturbation expansion can be seen as a violation of
identities of the form
$\langle \cdot | c^{\dagger}_{k\sigma'}\cdots c_{i\sigma }
c_{i \sigma }\cdots c_{j\sigma''}|\cdot \rangle=0$, when not all Wick contractions are evaluated. Such
expectation values will correctly vanish when the effective
interaction is used because the prefactor of the $c_{i\sigma}c_{i\sigma}$
operator, $\tilde V_{i\sigma,i\sigma}$, is zero. 
The presence of self-interaction errors in (non-self
consistent) $GW$ calculations was recently studied for a hydrogen
atom~\cite{selfint_H}.

In App. \ref{app.effint} we compare the performance of the effective interaction with
exact results for the Hartree and exchange self-energies of a benzene
molecule. These first order results indicate that the accuracy of $GW$ calculations
based on the effective interaction (\ref{eq.intapprox}) should be comparable to $GW$ calculations
based on the full interaction (\ref{eq.interaction}).
We stress, however, that in practice only the
correlation part of the $GW$ self-energy (second- and higher order
terms) is evaluated using $\hat
V_{\text{eff}}$, while the
Hartree and exchange self-energies are treated separately at a higher
level of accuracy, see Sec. \ref{sec.hartree_exchange}.

\subsection{Non-equilibrium $GW$ self-energy}\label{sec.noneqgw}
\begin{figure}[!h]
\includegraphics[width=0.98\linewidth]{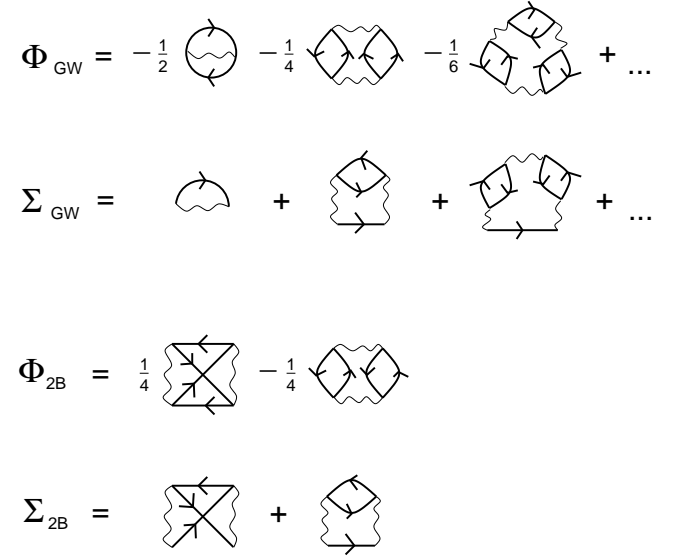}
\caption[cap.wavefct]{\label{fig.diagrams} The $GW$ and second Born
  self-energies, $\Sigma_{GW}$ and $\Sigma_{2B}$, can be obtained as
  functional derivatives of their respective $\Phi$-functionals,
  $\Phi_{GW}[G]$ and $\Phi_{2B}[G]$. Straight lines represent the full Green's
  function, $G$, i.e. the Green's function in the presence of coupling to the leads and
  interactions. Wiggly lines represent the interactions.}
\end{figure}

It is useful to split the full interaction self-energy into its Hartree and exchange-correlation parts
\begin{equation}
\Sigma(\tau,\tau')=\Sigma_h(\tau,\tau')+\Sigma_{xc}(\tau,\tau').
\end{equation} 
The Hartree term is local in
time and can be written
$\Sigma_h(\tau,\tau')=\Sigma_h(\tau)\delta_{\mathcal C}(\tau,\tau')$
where $\delta_{\mathcal C}$ is a delta function on the Keldysh
contour. Within the \emph{GW} approximation the exchange-correlation
term is written as a product of the
Green's function, $G$, and the screened interaction, $W$, calculated in the random-phase
approximation (RPA). With the effective interaction
(\ref{eq.intapprox}) the screened interaction and the polarization are
reduced from four- to two-index functions. For
notational simplicity we absorb the spin index into the orbital
index, i.e. $(i\sigma)\to i$ (but we do not neglect it). The $GW$ equations on the contour then
read
\begin{widetext}
\begin{eqnarray}\label{eq.gw_contour}
  \Sigma_{GW,ij}(\tau,\tau')&=&iG_{ij}(\tau,\tau'^+)W_{ij}(\tau,\tau')\\ 
  W_{ij}(\tau,\tau')&=&\tilde V_{ij}\delta_{\mathcal C}(\tau,\tau')+\sum_{kl}\int_{\mathcal C} \text{d}\tau_1 \tilde V_{ik}P_{kl}(\tau,\tau_1)W_{lj}(\tau_1,\tau')\label{eq.gw_contour2}\\
  P_{ij}(\tau,\tau')&=&-iG_{ij}(\tau,\tau')G_{ji}(\tau',\tau)\label{eq.gw_contour3}.
\end{eqnarray}
\end{widetext}
It is important to notice that in contrast to the conventional
real-space formulation of the $GW$ method, the spin-dependence cannot
be neglected when the effective interaction is used. The
reason for this is that $\tilde V$ is spin-dependent and consequently
the spin off-diagonal elements of $W$ will influence the spin-diagonal
elements of $G,\Sigma$, and $P$. A diagrammatic representation of the
$GW$ approximation is shown in Fig.~\ref{fig.diagrams}.

As they stand, equations (\ref{eq.gw_contour})-(\ref{eq.gw_contour3}) involve
quantities of the whole system (leads and central region). However, since $\tilde V_{ij}$
is non-zero only when $i,j\in C$, it follows from Eq.~(\ref{eq.gw_contour2}), that $W$ and hence $\Sigma$ also have
this structure. Consequently, the subscript $C$ can be directly attached to each
quantity in Eqs.~(\ref{eq.gw_contour})-(\ref{eq.gw_contour3}), however, for the
sake of generality and notational simplicity we shall not do so at this point.
It is, however, important to realize that the GF appearing in the $GW$
equations includes the self-energy due to the leads. 

Using the Langreth conversion rules~\cite{haug_jauho} the retarded and
lesser $GW$ self-energies become (on the time axis),
\begin{eqnarray}\label{eq.sigmar}
\Sigma^r_{GW,ij}(t)&=&iG^{r}_{ij}(t)W^{>}_{ij}(t)+iG^{<}_{ij}(t)W^{r}_{ij}(t)\\ \label{eq.sigmalg}
\Sigma^{</>}_{GW,ij}(t)&=&iG^{</>}_{ij}(t)W^{</>}_{ij}(t),
\end{eqnarray}
where we have used the variable $t$ instead of the time difference $t'-t$.
For the screened interaction we obtain (in frequency space),
\begin{eqnarray}\label{eq.wr}
W^{r}(\omega)&=&\tilde V[I-P^{r}(\omega)\tilde V]^{-1}\\\label{eq.wlg}
W^{</>}(\omega)&=&W^{r}(\omega)P^{</>}(\omega)W^{a}(\omega).
\end{eqnarray}
where all quantities are matrices in the indices $i,\sigma$ and matrix
multiplication is implied.  Notice that the spin off-diagonal part
of $\tilde V$ will affect the spin-diagonal part of $W^r$ through the
matrix inversion. 

Finally, the real-time components of the irreducible polarization become
\begin{eqnarray}\label{eq.pr}
P^{r}_{ij}(t)&=&-iG^{r}_{ij}(t)G^{<}_{ji}(-t)-iG^{<}_{ij}(t)G^{a}_{ji}(-t)\\\label{eq.pl}
P^{</>}_{ij}(t)&=&-iG^{</>}_{ij}(t)G^{>/<}_{ji}(-t).
\end{eqnarray}
From their definitions it is clear that both the polarization and the
screened interaction obey the relations
$P^{a}_{ij}(\omega)=P^{r}_{ji}(-\omega)$ and
$W^{a}_{ij}(\omega)=W^{r}_{ji}(-\omega)$, while for the self-energy and GFs we have $\Sigma_{GW}^a(\omega)=\Sigma_{GW}^r(\omega)^{\dagger}$ and $G^a(\omega)=G^r(\omega)^{\dagger}$. In addition all quantities
fullfill the general identity $X^>-X^<=X^r-X^a$. Notice that the GFs
entering the $GW$ equations are the We mention that
equations similar to those derived above without the extra
complication of coupling to external leads, have previously been used
to calculate bulk bandstructures of
excited GaAs.~\cite{louie04}

In deriving Eqs.~(\ref{eq.wr},\ref{eq.wlg}) we have
made use of the conversion rules $\delta_{\mathcal
  C}^{</>}(t,t')=0$ and $\delta_{\mathcal
  C}^{r/a}(t,t')=\delta(t-t')$. With these definitions the
applicability of Langeth rules can be extended to functions containing
delta functions on the contour. Notice, however, that with these
definitions relation (\ref{eq.ret}) does not hold for the delta
function. The reason why the delta function requires a separate
treatment is that the Langreth rules are derived under the assumption
that all functions on the contour are well behaved, e.g. not containing
delta functions.

We stress that no spin symmetry has been assumed in the above $GW$ equations. Indeed by
reintroducing the spin index, i.e. $i\to(i\sigma)$ and $j\to(j\sigma')$, it is clear that
spin-polarized calculations can be performed by treating $G_{\uparrow
  \uparrow}$ and $G_{\downarrow
  \downarrow}$ independently.
 
Within the $GW$ approximation the full interaction self-energy is
given by
\begin{equation}
\Sigma(\tau,\tau')=\Sigma_h(\tau,\tau')+\Sigma_{GW}(\tau,\tau'),
\end{equation}
where the $GW$ self-energy can be further split into an exchange and
correlation part,
\begin{equation}
\Sigma_{GW}(\tau,\tau')=\Sigma_x\delta_{\mathcal C}(\tau,\tau')+\Sigma_{\text{corr}}(\tau,\tau').
\end{equation}
Due to the static nature of $\Sigma_h$ and $\Sigma_x$ we have
\begin{equation}\label{eq.exlg}
\Sigma_h^{</>}=\Sigma_x^{</>}=0.
\end{equation}
The retarded components of the Hartree and exchange self-energies
become constant in frequency space, and we have (note that for
$\Sigma_h$ and $\Sigma_x$ we do \emph{not}
use the effective interaction (\ref{eq.intapprox}))
\begin{eqnarray}\label{eq.hr}
\Sigma^r_{h,ij}&=&-i\sum_{kl}G^<_{kl}(t=0) V_{ik,jl}\\ \label{eq.exr}
\Sigma^r_{x,ij}&=&i\sum_{kl}G^<_{kl}(t=0) V_{ik,lj}.
\end{eqnarray}
Due to (\ref{eq.exlg}), it is clear that Eq.~(\ref{eq.sigmalg}) yields
the lesser/greater components of $\Sigma_{\text{corr}}$. Since
$\Sigma_{\text{corr}}(\tau,\tau')$ does not contain
delta functions its retarded component can be obtained from the
relation,
\begin{equation}\label{eq.corrr}
\Sigma_{\text{corr}}^r(t)=\theta(-t)[\Sigma^>_{GW}(t)-\Sigma^<_{GW}(t)].
\end{equation}
The separate calculation of $\Sigma_x^r$ and $\Sigma_{\text{corr}}^r$ from
Eqs.~(\ref{eq.exr}),(\ref{eq.corrr}) as opposed to calculating their
sum directly from Eq.~(\ref{eq.sigmar}), has two advantages: (i) It allows us to treat
$\Sigma_x$, which is the dominant contribution to $\Sigma_{GW}$, at a
higher level of accuracy than $\Sigma_{\text{corr}}$, see
App.~\ref{app.hartree_exchange}. (ii) We avoid numerical
operations involving $G^r$ and $W^r$ in the
time domain, see App.~\ref{sec.retfromcorr}.

\subsection{Non-equilibrium second Born approximation} 
When screening and/or strong correlation effects are less important, as
e.g. in the case of small molecules, the higher-order terms of the $GW$
approximation are small and it is more important to include
all second order diagrams~\cite{leeuwen06}. The full second order
approximation, often referred to as the
second Born approximation (2B), is shown diagrammatically in
Fig.~\ref{fig.diagrams}. As we will use the 2B for comparison with the $GW$ results we state the relevant
expressions here for completeness. The non-equilibrium 2B has recently
been applied to study atoms in laser fields~\cite{leeuwen07}.

On the contour the 2B self-energy reads (with the effective interaction (\ref{eq.intapprox})) 
\begin{eqnarray}\nonumber
\Sigma_{2B,ij}(\tau,\tau')&=&\sum_{kl}G_{ij}(\tau,\tau')G_{kl}(\tau,\tau')G_{lk}(\tau',\tau)\tilde
V_{ik}\tilde
V_{jl}\\ \nonumber &-&\sum_{kl}G_{ik}(\tau,\tau')G_{kl}(\tau',\tau)G_{lj}(\tau,\tau')\tilde
V_{il}\tilde V_{jk}\\\label{eq.sigma2b}
\end{eqnarray}
Notice that the first term in $\Sigma_{2B}$ is simply the second order
term of the $GW$ self-energy. From Eq.~(\ref{eq.sigma2b}) it is easy to obtain the lesser/greater
self-energies,
\begin{eqnarray}\nonumber
\Sigma_{2B,ij}^{</>}(t)&=&\sum_{kl}G^{</>}_{ij}(t)G^{</>}_{kl}(t)G^{>/<}_{lk}(-t)\tilde
V_{ik}\tilde
V_{jl}\\ \nonumber &-&\sum_{kl}G^{</>}_{ik}(t)G^{>/<}_{kl}(-t)G^{</>}_{lj}(t)\tilde
V_{il}\tilde V_{jk},
\end{eqnarray}  
where $t$ has been used instead of the time difference $t-t'$. Since
these second order contributions do not contain delta functions of
the time variable, we can obtain the retarded self-energy directly
from the Kramers-Kronig relation
\begin{equation}
\Sigma_{2B}^r(t)=\theta(-t)[\Sigma_{2B}^>(t)-\Sigma_{2B}^<(t)],
\end{equation}
see App.~\ref{sec.retfromcorr}.

\section{Current Formula and Charge conservation}\label{sec.conserving}
In this section we address the question of charge conservation in the
model introduced in Sec.~\ref{sec.model}. In particular, we ask under
which conditions the current calculated at the left and right sides of
the central region are equal, and we show in Sec.~\ref{sec.consfromphi} that this is fulfilled whenever
the self-energy used to describe the interactions is $\Phi$-derivable, independently of the applied basis set.

\subsection{Current formula}\label{sec.current}
As shown by Meir and Wingreen~\cite{meir_wingreen92}, the particle current 
  from lead $\alpha$ into the central region can be expressed as
\begin{equation}\label{eq.current2}
I_{\alpha}=\int \frac{\text{d} \omega}{2\pi}\text{Tr}\big [
\Sigma_{\alpha}^<(\omega) G_C^>(\omega)-\Sigma_{\alpha}^>(\omega) G_C^<(\omega)\big ],
\end{equation}
where matrix multiplication is understood. By 
writing $I=(I_L-I_R)/2$ one obtains a current expression symmetric in the $L,R$ indices, 
\begin{equation}\label{eq.symcurrent}
I=\frac{i}{4\pi}\int \text{Tr}[(\Gamma_L-\Gamma_R)G_C^{<}+(f_L\Gamma_L-f_R\Gamma_R)(G_C^r-G_C^a)]\text{d}\omega
\end{equation}
where we have suppressed the $\omega$ dependence and introduced the
coupling strength of lead $\alpha$,
$\Gamma_{\alpha}=i[\Sigma^r_{\alpha}-\Sigma^a_{\alpha}]$. We note in
passing that for non-interacting electrons the integral has weight
only inside the bias window whereas this is no longer true when
interactions are present.

\subsection{Charge conservation}\label{sec.currentcons}
Due to charge conservation we expect that in steady-state 
$I_L=-I_R=I$, i.e. the current flowing from the left lead to the
molecule is the negative of the current flowing from the right lead to
the molecule. Below we derive a condition for this specific
form of particle conservation.

From Eq.~(\ref{eq.current2}) the difference between the currents at the left and right interface, $\Delta I=I_L+I_R$, is given by
\begin{equation}\label{eq.currentvanishcond2}
\Delta I=\int \frac{\text{d} \omega}{2\pi}\text{Tr}\big [
(\Sigma^<_L+\Sigma^<_R) G_C^>-(\Sigma^>_L+\Sigma_R^>) G_C^<\big ]
\end{equation}
To obtain a condition for $\Delta I=0$ in terms of $\Sigma$ we start by proving the general identity 
\begin{equation}\label{eq.currentvanishcond}
\int \frac{\text{d} \omega}{2\pi}\text{Tr}\big [
\Sigma_{tot}^<(\omega) G_C^>(\omega)-\Sigma_{tot}^>(\omega)G_C^<(\omega)\big ]=0.
\end{equation}
To prove this, we insert $G^{</>}=G^r_C
\Sigma_{tot}^{</>}G^a_C+\Delta^{</>}$ (from Eq.~(\ref{eq.keldysh})) in 
the left hand side  of Eq.~(\ref{eq.currentvanishcond}). This results in two
terms involving $G^r \Sigma_{tot}^{</>}G^a$ and two terms involving
$\Delta^{</>}$. The first two terms contribute by 
\begin{equation}\label{eq.keldyshvanish}
\int \frac{\text{d} \omega}{2\pi}\text{Tr}\big [
\Sigma_{tot}^< G^r \Sigma_{tot}^> G^a -\Sigma_{tot}^>G^r \Sigma_{tot}^< G^a\big ].
\end{equation}
Inserting $\Sigma^>_{tot}=\Sigma_{tot}^<+(G^a)^{-1}-(G^r)^{-1}$ (see
note \onlinecite{help1}) in this expression and using the cyclic invariance of the trace, it is straightforward to show that Eq.~(\ref{eq.keldyshvanish}) vanishes.
The two terms involving $\Delta^{</>}$ contribute to the left hand side of Eq.~(\ref{eq.currentvanishcond}) by
\begin{equation}\label{eq.deltavanish}
\int \frac{\text{d} \omega}{2\pi}\text{Tr}\big [
\Sigma^<_{tot}(\omega) \Delta^>(\omega)-\Sigma^>_{tot}(\omega)\Delta^<(\omega)\big ].
\end{equation}
As discussed in Sec.~(\ref{sec.boundstates}) $\Delta^{<}$ and
$\Delta^{>}$ are always zero when interactions are present. In the
case of non-interacting electrons we have
$\Sigma^{</>}_{tot}=\Sigma_L^{</>}+\Sigma_R^{</>}$, which vanish
outside the band width the leads. On the other hand $\Delta^{</>}$ is
only non-zero at energies corresponding to bound states, i.e. states
lying outside the bands, and thus we conclude that the term~(\ref{eq.deltavanish}) is always zero.

From Eqs.~(\ref{eq.currentvanishcond2}) and
(\ref{eq.currentvanishcond}) it then follows that
\begin{equation}\label{eq.currentcondition}
\Delta I=\int \frac{\text{d} \omega}{2\pi}\text{Tr}\big [
\Sigma^<(\omega) G_C^>(\omega)-\Sigma^>(\omega)G_C^<(\omega)\big ].
\end{equation}
We notice that without any interactions particle conservation in the
sense $\Delta I=0$ is trivially fulfilled since $\Sigma=0$. When
interactions are present, particle conservation depends on the
specific approximation used for the interaction self-energy, $\Sigma$. 

\subsection{Conserving approximations}
A self-energy is called conserving, or $\Phi$-derivable, 
if it can be written as a functional derivative of a so-called $\Phi$-functional,
$\Sigma[G]=\delta \Phi[G]/\delta G$.\cite{baym62} Since a $\Phi$-derivable self-energy depends on $G$, the Dyson equation must be solved
self-consistently. The resulting Green's function automatically fulfills all important conservation
laws including the continuity equation which is of major relevance
the context of quantum transport. 

The exact $\Phi[G]$ can be
obtained by summing over all skeleton diagrams, i.e. closed diagrams
with no self-energy insertions, constructed using the
full $G$ as propagator. Practical approximations are then obtained by
including only a subset of skeleton diagrams. Two examples of such 
approximations are provided by the $GW$ and second Born $\Phi$-functional and associated
self-energies which are illustrated in Fig.~\ref{fig.diagrams}. Solving the
Dyson equation self-consistently with one of these self-energies thus defines a
conserving approximation in the sense of Baym.

The validity of the conservation laws for $\Phi$-derivable
self-energies follows from the invariance of $\Phi$ under certain
transformations of the Green's function. For example it follows from
the closed diagramatic structure of $\Phi$ that the
transformation~\cite{baym62}
\begin{equation}\label{eq.Gtrans}
G(\bold r \tau,\bold r' \tau')\to e^{i\Lambda(\bold r \tau)}G(\bold r \tau,\bold r' \tau')e^{-i\Lambda(\bold r' \tau')},
\end{equation}
where $\Lambda$ is any scalar function, leaves $\Phi[G]$ unchanged.
Using the compact notation $(\bold r_1,\tau_1)=1$, the change in
$\Phi$ when the GF is changed by $\delta G$ can be written as
$\delta\Phi=\int \text{d}1 \text{d} 2 \Sigma(1,2)\delta G(2,1^+)=0$,
where we have used that $\Sigma=\delta \Phi[G]/\delta G$. To first
order in $\Lambda$ we then have
\begin{eqnarray}\nonumber
\delta\Phi&=&i\int \text{d}1 \text{d}2
\Sigma(1,2)[\Lambda(2)-\Lambda(1)]G(2,1^+)\\ \nonumber
&=&i\int \text{d}1 \text{d}2 [\Sigma(1,2)G(2,1^+)-G(1,2^+)\Sigma(2,1)]\Lambda(1).
\end{eqnarray}
Since this hold for all $\Lambda$ (by a scaling argument) we conclude that 
\begin{equation}\label{eq.realspacecond}
\int \text{d}2 [\Sigma(1,2)G(2,1^+)-G(1,2^+)\Sigma(2,1)]=0.
\end{equation}
It can be shown that this condition ensures the validity of the continuity equation
(on the contour) at any point in space~\cite{baym62}.

\subsection{Charge conservation from $\Phi$-derivable
  self-energies}\label{sec.consfromphi}
Below we show that $\Delta I$ of Eq.~(\ref{eq.currentcondition})
always vanishes when the self-energy is $\Phi$-derivable, i.e. the
general concept of a conserving approximation carries over to the
discrete framework of our transport model. 

We start by noting that Eq.~(\ref{eq.realspacecond})
holds for \emph{any} pair $G(1,2),\Sigma[G(1,2)]$ provided $\Sigma$ is
of the $\Phi$-derivable form. In particular
Eq.~(\ref{eq.realspacecond}) does not assume that the pair $G,\Sigma[G]$
fulfill a Dyson equation. Therefore, by taking any orthonormal, but not
necessarily complete set, $\{\phi_i\}$, and writing
$G(1,2)=\sum_{ij}\phi_i(\bold r_1)G_{ij}(\tau_1,\tau_2)\phi_j^*(\bold
r_2)$ we get from Eq.~(\ref{eq.realspacecond}) after integrating over
$\bold r_1$,
\begin{equation}\label{eq.matrixcond0}
\sum_j \int_{\mathcal C} \text{d}\tau' [\Sigma_{ij}(\tau,\tau')G_{ji}(\tau',\tau^+)-G_{ij}(\tau^-,\tau')\Sigma_{ji}(\tau',\tau)]=0,
\end{equation}
which in matrix notation takes the form
\begin{equation}\label{eq.matrixcond}
\int_{\mathcal C} \text{d}\tau' \text{Tr}[\Sigma(\tau,\tau')G(\tau',\tau^+)-G(\tau^-,\tau')\Sigma(\tau',\tau)]=0.
\end{equation}
Here $\Sigma_{ij}$ is exactly the self-energy matrix obtained when the
diagrams are evaluated using $G_{ij}$ and the $V_{ij,kl}$ from
Eq.~(\ref{eq.interaction}).  
The left hand side of Eq.~(\ref{eq.matrixcond}), which is
always zero for a $\Phi$-derivable $\Sigma$, can be written as
$\text{Tr}[A^<(t,t)]$ when $A$ is given by Eq.~(\ref{app.defA}) with
$B=\Sigma$ and $C=G$. It then follows from the general result (\ref{app.result}) and the condition (\ref{eq.currentcondition}) that current
conservation in the sense $I_L=-I_R$ is always obeyed when $\Sigma$ is
$\Phi$-derivable.

The above derivation of Eq.~(\ref{eq.matrixcond}) relied on \emph{all}
the Coulomb matrix elements, $V_{ijkl}$, being included in the
evaluation of $\Sigma$. Thus the proof does not carry through if a
general truncation scheme for the interaction matrix is used.
However, in the special case of a truncated interaction of the form
(\ref{eq.intapprox}), i.e. when the interaction is a two-point
function, Eq.~(\ref{eq.matrixcond}) remains valid. To show this, it is
more appropriate to work entirely in the matrix representation and
thus define $\Phi[G_{ij}(\tau,\tau')]$ as the sum of a set of skeleton
diagrams evaluated directly in terms $G_{ij}$ and $\tilde
V_{ij}$. With the same argument as used in Eq.~(\ref{eq.Gtrans}), it
follows that $\Phi$ is invariant under the transformation
\begin{equation}
G_{ij}(\tau,\tau')\to e^{i\Lambda_i(\tau)}G_{ij}(\tau,\tau')e^{-i\Lambda_j(\tau')},
\end{equation}
where $\Lambda$ is now a discrete vector. By adapting the arguments
following Eq.~(\ref{eq.Gtrans}) to the discrete case we arrive at
Eq.~(\ref{eq.realspacecond}) with the replacements $\bold r_1 \to i$
and $\bold r_2 \to j$ and with the integral replaced by a discrete sum
over $j$. Summing also over $i$ leads directly to
Eq.~(\ref{eq.matrixcond}) which is the desired result.

To summarize, we have shown that particle conservation in the sense $I_L=-I_R$, is
obeyed whenever a $\Phi$-derivable self-energy is used \emph{and} either (i) all
Coulomb matrix elements $V_{ij,kl}$ or (ii) the truncated
two-point interaction of Eq.~(\ref{eq.intapprox}), are used to evaluate $\Sigma$.

\section{Implementation}\label{sec.method}
In this section we describe the practical implementation of the
Wannier-$GW$ transport scheme. After a brief sketch of the basic idea of the method we
outline the calculation of the non-interacting Hamiltonian matrix
elements and Coulomb
integrals in terms of Wannier orbitals. The explicit expression for the Green's function is given in
Sec.~\ref{sec.exprGr}, and in Sec.~\ref{sec.selfconsistent} we
describe our implementation of the Pulay mixing scheme for performing
self-consistent Green's function calculations. We end the section with a discussion
of the present limitations and future improvements of the method.

\subsection{Interactions in the central region}
Most first-principles calculations addressing transport in molecular
contacts are based on the assumption that the charge carriers (electrons) can be
  considered as independent particles governed by an effective
  single-particle Hamiltonian. A popular choice for the effective Hamiltonian is the Kohn-Sham Hamiltonian of DFT,
\begin{equation}
\hat h_{s}=-\frac{1}{2}\nabla^2+v_{ext}(\bold r)+v_h(\bold r)+v_{xc}(\bold r),
\end{equation} 
where $v_{ext}(\bold r)$ is the external potential from the ions,
$v_h(\bold r)$ is the classical Hartree field, and $v_{xc}(\bold r)$
is the exchange-correlation (xc-) potential which to some degree
includes e-e interaction effect beyond the Hartree level. 

In the present method we rely on the KS Hamiltonian to describe the
metallic electrodes as well as the coupling into the central region,
but replace the local xc-potential by a many-body 
self-energy inside the central region where correlation effects are
expected to be most important. Clearly, this division does not
treat all parts of the system on the same footing, and one might be
concerned that electrons can scatter off the artificial interface
defined by the transition region between the mean-field and many-body
description and thus introduce an artificial "contact resistance".
Such unphysical scattering is certainly expected to affect the
calculated properties if the transition region is very close to the
constriction of the contact. On the other hand, the central region
can, at least in principle, be chosen so large that the transition
region occurs deep in the electrodes far away from the constriction.
In this case the large number of available conductance channels in the
electrodes should ensure that the calculated properties are not
dominated by interface effects and the non-interacting part of the
electrodes will mainly serve as particle reservoirs whose precise
structure is unimportant. Thus the assumption of
interactions in the central region seems justified in principle
although it might be difficult to fully avoid artificial backscattering in practice.

\subsection{Wannier Hamiltonian and Coulomb integrals}\label{sec.basis}
In order to make the evaluation and storing of the $GW$ self-energy
feasible, we use a minimal basis set consisting of maximally
localized, partially occupied Wannier
functions~\cite{WFprb} obtained from the plane-wave pseudopotential
code Dacapo~\cite{dacapo}. Below we outline how the Hamiltonian is
evaluated in the WF basis and refer to Ref.~\onlinecite{thygesen_chemphys} for more details.

\begin{figure}[!b]
\includegraphics[width=0.95\linewidth]{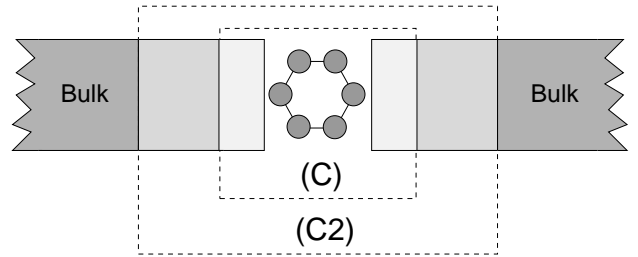}
\caption[cap.wavefct]{\label{fig.gwregion} The extended central region
($C2$) is chosen so large that it comprises all perturbations in the
effective DFT potential arising from the molecular contact. The central
region ($C$) can be a proper subregion of $C2$, but it must be so large that there is no direct coupling
across it. We solve for the self-consistent Kohn-Sham potential within
$C2$, but replace the static xc potential by the $GW$ self-energy inside $C$.}
\end{figure}

The WFs used to describe the leads are obtained from a bulk
calculation (or supercell calculation if the leads have finite cross
section). We define the \emph{extended central region} ($C2$) as the
molecule itself plus a portion of the leads. $C2$ should be so large
that it comprises all perturbations in the KS potential arising from
the presence of the molecular contact such that a smooth transition
from $C2$ into the bulk is ensured. The WFs inside $C2$ are obtained
from a DFT calculation with periodic boundary conditions imposed on
the supercell containing $C2$. The resulting WFs will inherit the
periodicity of the eigenstates, however, due to their localized nature
they can be unamigously extended into the lead regions.  Thanks to the
large size of $C2$, hybridization effects between the molecule and the
metal leads will automatically be incorporated into the WFs. With the
combined set of WFs (lead+$C2$), we can then represent any KS state of
the contacted system up to a few electron volts above the Fermi energy.~\cite{WFprb}.

In practice, the requirement of complete screening means that 3-4
atomic layers of the lead material must be included in $C2$ on both
sides on the molecule. While this size of systems can be easily
handled within DFT it may well exceed what is computationally feasible
for a many-body treatment such as the $GW$ method even with the
minimal WF basis. For this reason we shall allow the central region
($C$) to consist of a proper subset of the WFs in $C2$, subject to the
requirement that there is no direct coupling across it, i.e. $\langle
\phi_i|\hat h_s|\phi_j\rangle=0$ for $i\in L$ and $j\in R$ where the
left (right) lead by definition is all WFs to the left (right) of
$C$. With this definition of $C$, the KS potential outside $C$ is not
necessarily periodic (this is, however, always the case outside $C2$),
and consequently the calculation of the coupling self-energies becomes
somewhat more involved as compared to the usual situation of periodic
leads, see discussion in App.~\ref{sec.aperleads}. We stress that the
transmission function for the non-interacting KS problem is
exactly the same whether $C$ or $C2$ is used as the central region as
long as there is no direct coupling across region $C$.

Having constructed the WFs we calculate the matrix elements of the
effective KS Hamiltonian of the \emph{contacted, unbiased} system, $\langle \phi_i | \hat
h_{s}|\phi_j\rangle$. 
To correct for double counting when the $GW$ self-energy is added, we also need the matrix elements, $\langle \phi_i |
v_{xc}|\phi_j\rangle$, for WFs belonging to the central region. 

The matrix elements defining the interaction 
$\hat V$ in Eq.~(\ref{eq.interaction}) are calculated as the
(unscreened) Coulomb integrals
\begin{equation}\label{eq.coulombints}
V_{ij,kl}=\int \int \text{d}\bold r \text{d}\bold r' \frac{\phi_i(\bold r)^*\phi_j(\bold r')^* \phi_k(\bold r)\phi_l(\bold r') }{|\bold r - \bold r'|},
\end{equation}
for WFs belonging to the central region. The Coulomb
integrals are evaluated in Fourier space using neutralizing Gaussian
charge distributions to avoid contributions from the periodic images,
see note \onlinecite{gauss}. 

\subsection{Hartree and exchange}\label{sec.hartree_exchange}
As already mentioned it is not feasible
to include all the interaction matrix elements when
evaluating the frequency-dependent part of the many-body self-energy,
$\Sigma_{\text{corr}}$, which is therefore calculated using effective interaction of Eq.~(\ref{eq.intapprox}). 

However, the exchange term, which can be unambiguously separated from the $GW$
self-energy, is evaluated from Eq.~(\ref{eq.exr}) using all Coulomb
elements of the forms
$\{\{V_{ij,ij}\},\{V_{ij,ji}\},\{V_{ii,jj}\},\{V_{ii,ij}\}\}$. As
shown in appendix \ref{app.hartree_exchange} this produces results within $5\%$
of the exact values.

The KS Hamiltonian already includes the Hartree potential of the DFT
groundstate. In a self-consistent, finite bias $GW$ calculation the
relevant Hartree potential will deviate from the DFT Hartree potential
due to the finite bias and the fact that the xc-potential is replaced
by the $GW$ self-energy. This correction, which is much smaller than the full
Hartree potential, is treated in the same way as the exchange term,
i.e. calculated from Eq.~(\ref{eq.exr}) with all Coulomb
elements of the form
$\{\{V_{ij,ij}\},\{V_{ij,ji}\},\{V_{ii,jj}\},\{V_{ii,ij}\}\}$. As for
the exchange terms this yields results within $5\%$
of the exact values, see \ref{app.hartree_exchange}.

\subsection{Expression for $G^r$}\label{sec.exprGr}
To simplify the notation in the following we omit the subscript $C$ as all quantities will be matrices in the central region. The retarded GF of the central region is obtained from
\begin{widetext}
\begin{equation}\label{eq.grfromdft}
G^r=[(\omega+i\eta)I-(h_{s}-v_{xc})-\Sigma^r_L-\Sigma^r_R-(\Sigma^r_h[G]-\Sigma^r_h[g_{s}^{\text{(eq)}}])-\Sigma^r_{GW}[G]]^{-1}
\end{equation}
\end{widetext}
Several comments are in order. First, we notice that all
quantities except for $v_{xc}$, $h_s$, and
$\Sigma^r_h[g_{s}^{\text{(eq)}}]$, are bias-dependent, however, to keep the notation as
simple as possible we omit any reference to this dependence. The terms
$\Sigma^r_L$ and $\Sigma^r_R$ account for the coupling to the leads. By
subtracting $v_{xc}$ from $h_s$ we ensure that exchange-correlation
effects are not counted twice when we add the $GW$ self-energy,
$\Sigma^r_{GW}$. The term
$\Delta v_h=\Sigma^r_h[G]-\Sigma^r_h[g_{s}^{\text{(eq)}}]$ is the change in
Hartree potential relative to the equilibrium DFT value. This change
is due to the
applied bias and the replacement of $v_{xc}$ by $\Sigma^r_{GW}$ (even in equilibrium the Hartree field will
change during the $GW$ self-consistency cycle). The Hartree potential
in $C$ originating from the electron density in the electrodes, which enters
$G^r$ through $h_s$, is assumed to
stay constant when the system is driven out of equilibrium, i.e. the
out-of-equilibrium charge distribution in the leads is assumed to equal the equilibrium one.

Finally, in order to make contact with the general formalism of
Sec.~(\ref{sec.formalism}), and in particular
Eq.~(\ref{eq.grc_closed}), we note that the matrix elements $h_{ij}$
defining the effective single-particle Hamiltonian in Eq.~(\ref{eq.nonintham}), are related to the quantities introduced above via
\begin{displaymath}\label{eq.hamiltonianmatrix}
h_{ij}=\left \{ \begin{array}{ll}
\langle \phi_i|\hat h_{s}-\hat
v_{xc}|\phi_j\rangle-\Sigma^r_h[g_{s}^{\text{(eq)}}]_{ij} &\text{ for $i,j$
  both in $C$}\vspace{2mm} \\
\langle \phi_i|\hat h_{s}|\phi_j\rangle+(\mu_{L(R)}-\varepsilon_{F})\delta_{ij} &\text{ for $i,j$
  both in $L(R)$}\vspace{2mm} \\
\langle \phi_i | \hat h_{s}|\phi_j\rangle & \text{ otherwise}
\end{array} \right.
\end{displaymath}

\subsection{Frequency dependence}
  To represent the temporal dependence of the Green's functions and
  $GW$ self-energies we use an equidistant frequency grid with $N_g$
  grid points and grid spacing $\delta$. Thus the GFs
  (and the $GW$ self-energies) are represented by $N_w\times N_w \times N_g$
  matrices. At each of the discrete frequencies $\omega_i=n_i \delta$,
  $n_i=0\ldots N_g$, we have an $N_w\times N_w$ matrix representation of $G(\omega_i)$
  in the WF basis.
  The grid spacing, $\delta$, should be small enough that all
  features in the frequency dependence of the GFs and self-energies can be
  resolved. At the same time the frequency grid should be large enough
  (contain enough points) to properly describe asymptotic behavior
  (the tail) of the GFs.
  Although the tail is irrelevant for the current in
  Eq.~(\ref{eq.symcurrent}), it contributes to the self-energy,
  $\Sigma_{GW}[G]$. In practice, $N_g$ and $\delta$ should be
  increased, respectively decreased, until the results do not
  change. 

To avoid time consuming convolutions on the
  frequency grid, we use the Fast Fourier Transform (FFT) to switch
  between frequency and time domains. An important but technical issue
  concerning the evaluation of retarded functions is discussed in
  App.~\ref{sec.retfromcorr}.

\subsection{Self-consistency}\label{sec.selfconsistent}
Since $\Sigma$ depends on $G$, and $G$ depends on $\Sigma$, the Dyson
equations Eqs.~(\ref{eq.keldysh}) and (\ref{eq.grfromdft}) must be solved
self-consistently in conjunction with the equations for the $GW$,
Hartree, and exchange self-energies. In practice this self-consistent
problem is solved by
iteration. Clearly, the iterative approach relies on the assumption
that the problem has a unique solution and that the iterative process converges to this
solution. For all applications we have studied so far this has been
the case. In order to stabilize the iterative procedure, we use the
Pulay scheme~\cite{pulay} to mix the GFs of the previous $N$
iterations very analogue to what is done for the electron density in
many DFT codes. More specifically the input
GF at iteration $n$ is obtained according to 
\begin{equation} 
G^{X,n}_{\text{in}}=(1-\alpha)\sum_{j=n-N}^{n-1}c_{j}^nG^{X,j}_{\text{in}}+\alpha
\sum_{j=n-N}^{n-1}c_{j}^nG^{X,j}_{\text{out}},\quad X=<,r
\end{equation}
To determine the optimal values for the expansion coefficients, $c^n$,
we first define an inner product in the space of (retarded) GFs
\begin{equation}
\langle G^{r,i},G^{r,j}\rangle=\sum_n\int
\text{Im}[G^{r,i}_{nn}(\omega)]^*\text{Im}[G_{nn}^{r,j}(\omega)]
\text d\omega.
\end{equation}
Equivalent inner products can be obtained e.g. by using the real part
of the GF instead of the imaginary part or the lesser component
instead of the retarded.
The Pulay residue matrix determining the coefficients $c^n$ is then given by 
\begin{equation}
A_{ij}^n=\langle
G^{r,i}_{\text{in}}-G^{r,i}_{\text{out}},G^{r,j}_{\text{in}}-G^{r,j}_{\text{out}}\rangle,
\end{equation}
where $i,j=n-N,\ldots,n-1$.
We typically use a mixing factor around $\alpha\approx 0.4$. 
During the mixing procedure one must keep track of both the retarded
and lesser GF since one does not follow directly from the other. However, it is
important that the \emph{same} coefficients, $c^n$, are used
for mixing the two components. If separate coefficients are used for
$G^r$ and $G^<$, the
fundamental relation (\ref{eq.fundrelation}) is not guaranteed
during the self-consistent cycle. As noted above we define the
residue exclusively from the retarded GF. In practice we always find
that once the retarded GF has converged, the lesser GF has converged
too, and this justifies the use of common expansion coefficients for
the two GF components.

\subsection{Overview}
Below we give an overview of the various steps involved in performing a
self-consistent non-equilibrium $GW$ transport calculation:

\begin{itemize}
\item Perform DFT calculations for the electrodes and the
extended central region (region $C2$ in Fig.~\ref{fig.gwregion}).

\item Construct the Wannier functions, and obtain the matrix
representation of the KS Hamiltonian for the contacted system in
equilibrium. Evaluate the matrix elements for $v_{xc}$ and
relevant Coulomb integrals for Wannier functions belonging to the
central region ($C$).

\item Fix the bias voltage, and calculate the coupling self-energies
Eq.~(\ref{eq.sigmacontour}) as described in
App.~\ref{sec.aperleads} (these stay unchanged during self-consistency).

\item Evaluate the initial (non-interacting) Green's functions,
$G^r_C$ and $G^<_C$, e.g. from the KS Hamiltonian.

\item From $G^r_C$ and $G^<_C$, construct the desired interaction self-energies ($\Sigma_h$,
$\Sigma_x$, $\Sigma_{GW}$, or $\Sigma_{2B}$).

\item Test for self-consistency. In the negative, obtain a new set of
output Green's functions from Eqs.~(\ref{eq.grfromdft}) and (\ref{eq.keldysh}) and
mix with the previous GFs as described in Sec.~\ref{sec.selfconsistent}. 
\end{itemize}

\subsection{Limitations and future improvements}
The main approximation of the present implementation is the use of a
fixed, minimal basis set. We have used WFs obtained from the DFT-PBE
orbitals, however, one could also use Hartree-Fock or some other mean-field orbitals. Out of equilibrium the
WFs will be distorted due to the change in electrostatic potential, however,
this effect is not included. Although
the manifold spanned by the WFs, i.e. the KS eigenstates up to a few
electron volts above the Fermi level, are expected to represent the GW
quasiparticle wave functions of the same energy range quite well, an
accurate representation of the screened interaction might require
inclusion of high-energy eigenstates.

With the present implementation of the $GW$ scheme it is not feasible
to include more than a few electrode atoms in addition to the molecule
itself in the $GW$ region (region $C$ in Fig.~\ref{fig.gwregion}).  
The use of a small central region region might affect the
description of image charge formations in the electrode, and it might
introduce artificial backscattering at the DFT-GW interface.

The use of larger and more accurate basis sets as well the
inclusion of more electrode atoms in the $GW$ region are not fundamental
but practical limitations of the method, which in principle could be removed by
invoking efficient simplifications/approximations into the present formalism.

\section{Anderson model}\label{sec.anderson}
Since its introduction in 1961 the Anderson impurity
model~\cite{anderson61} has become a standard tool to investigate strong correlation phenomena such as local moments formation, Kondo effects and
Coulomb blockade. The Anderson model describes a localized electronic level of energy
$\varepsilon_c$ and correlation energy $U$ coupled to a continuum of
states. Thus the central region-part of the Hamiltonian reads
\begin{equation}
\hat H_{C}=\varepsilon_c c^{\dagger}c +Un_{\uparrow}n_{\downarrow}.
\end{equation}

In equilibrium, accurate results for
the thermodynamic properties of the Anderson model have been obtained from the Bethe
ansatz~\cite{lowenstein83,wiegman83}, quantum monte carlo simulations~\cite{jarrel89,silver90},
and numerical renormalization group theory~\cite{wilson75,costi94}. 

Out of equilibrium, the low-temperature properties of the Anderson
model have been much less studied. The earliest work addressed the
problem by applying second-order perturbation theory in the
interaction strength $U$.\cite{hershfield91,yeyati93} Despite the
simplicity of this approach it provides a surprisingly good
description of the (equilibrium) spectral function. There are,
however, several fundamental problems related to the non-self
consistent low-order perturbative approach: (i) the result depends on
the starting point around which the perturbation is applied, (ii) it
inevitably violates the conservation laws, and (iii) it applies only
in the small-$U$ limit. Methods relying on the slave-boson technique~\cite{coleman84}
have been developed to explore the strong correlation regime of the
model. The noncrossing approximation is believed to work well in the
infinite-$U$ limit and for sufficiently small tunneling strength,
$\Gamma$, but it fails to reproduce the correct Fermi liquid behavior
at low temperatures.~\cite{meir_wingreen94,meir93} More recently, a
finite-$U$ slave-boson mean-field approach~\cite{dong01} has been
proposed.  Finally we mention that a number of
more advanced schemes have been used to address non-equilibrium
Kondo-like phenomea focusing on the low-energy properties of the
Anderson model in the limit where $U$ is much larger than the
hybridization energy, $\Gamma$~\cite{paaske04,schiller95,konik01} .
 
While the Anderson model is normally used to describe strongly correlated systems, the main application of the $GW$ approximation has
been to weakly interacting quasi-particles in
closed shell systems such as molecules, insulators and
semi-conductors. In view of this, one could argue that the $GW$
method is inappropriate for the Anderson model.
Nevertheless, we find this application rather instructive as it illustrates some
general features of the $GW$ approximation including the role of
self-consistency both in relation to
charge conservation and the line shape of spectral functions. 
Moreover, as many important transport
phenomena, like Kondo effects and Coulomb blockade, are well described
by the Anderson model, it should always be of interest to
benchmark a transport scheme against this model. 

In a very recent study~\cite{wang07}, the $GW$ approximation was
applied to the Anderson model in equilibrium for interaction strengths 
$U/\Gamma$ up to $8.4/0.65\approx 13$, and various temperatures. For
the largest interaction strength it was found that $GW$
prefers to break the spin symmetry leading
to directly erroneous results in the Kondo regime. For intermediate interaction strengths
($U/\Gamma=4.2/0.65\approx 6.5$) where $GW$ does not
break the spin symmetry, it was concluded that $GW$ does not describe
the $T$-dependence of the Kondo effect
well. Nevertheless, we show here that at $T=0$ the width of the $GW$
Kondo-like resonance follow the analytical result for $T_K$ quite
well for intermediate interaction strengths. 

Here, as in our previous paper~\cite{thygesen_gw}, we focus on the
zero temperature, non-equilibrium situation. We consider
interaction strengths of $U/\Gamma$ up to 8 (we keep fix $U=4$ and
vary $\Gamma$). For these interaction strengths we
always find a stable non-magnetic $GW$ solution, i.e. $G_{\uparrow
  \uparrow}=G_{\downarrow \downarrow}$. In contrast, the HF solution
can develop a magnetic moment for $U/\Gamma>\pi$ (depending on bias voltage
and $\varepsilon_c$). We adopt the wide-band approximation where the coupling to
the continuum is modeled by constant imaginary self-energies
$\Sigma_L+\Sigma_R=-i\Gamma$.  Without loss of generality we set
$E_F=0$. In all calculations the frequency grid extends from -15 to 15
with the grid spacing ranging from 0.1 to 0.0005.

\begin{figure}[!b]
\includegraphics[width=1.0\linewidth]{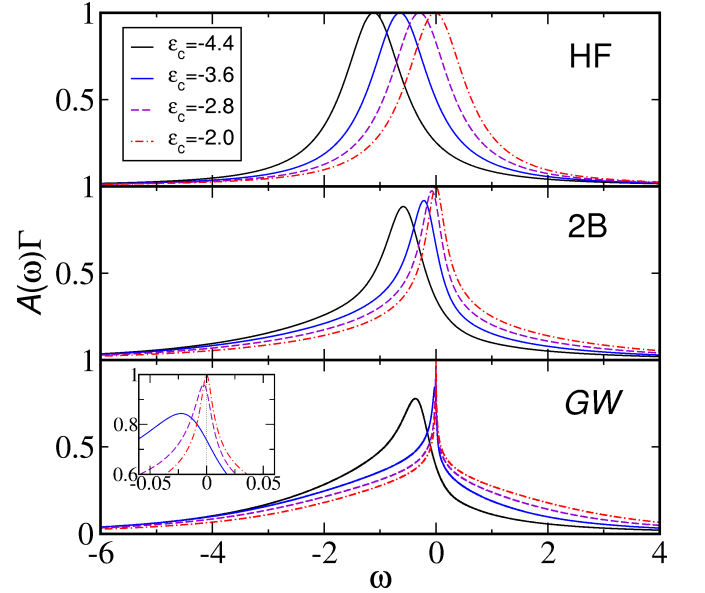}
\caption[cap.wavefct]{\label{fig.spec} (color online). Spectral
  function of the central site for
  $\Gamma=0.65$, $U=4.0$ and different values of $\varepsilon_c$. The inset in the lower panel is
  a zoom of the $GW$ spectral peak around $\omega=0$.}
\end{figure}

\begin{figure}[!b]
\includegraphics[width=0.9\linewidth,angle=270]{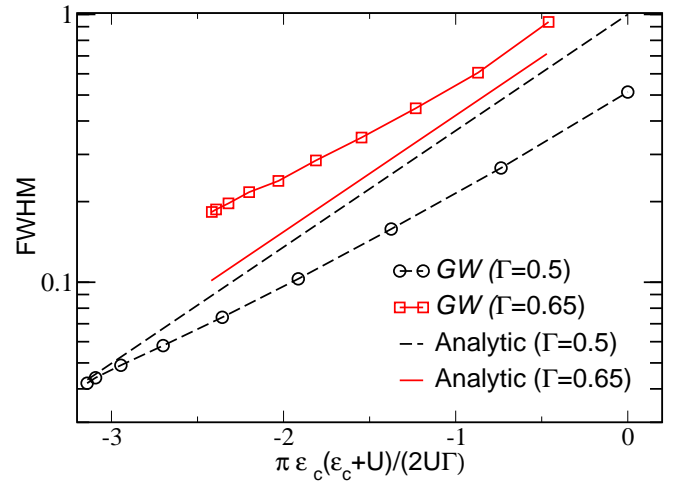}
\caption[cap.wavefct]{\label{fig.tk} (color online). Full width at
  half maximum (FWHM) of the Kondo resonance as calculated in the $GW$
approximation and from the analytical result Eq. (\ref{eq.tk}). The
  interaction strength is $U=4$ and $\varepsilon_c$ is varied in the Kondo regime.}
\end{figure}

\begin{figure}[!b]
\includegraphics[width=0.8\linewidth,angle=270]{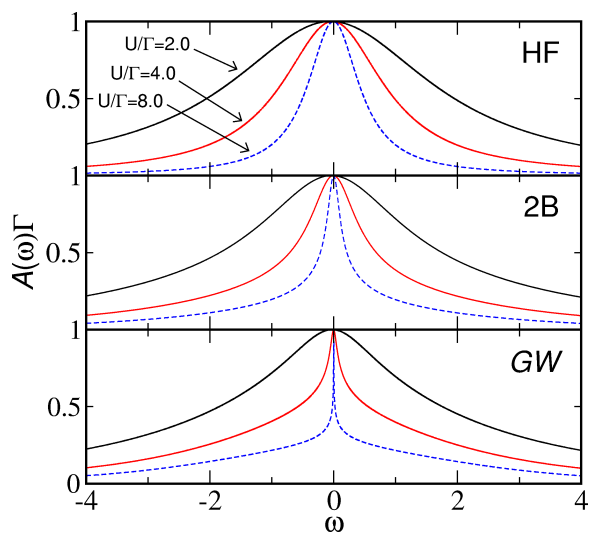}
\caption[cap.wavefct]{\label{fig.spec2} (color online). Spectral
  function for
  $U=4.0$, $\varepsilon_c=-U/2$, and three
  different values of $\Gamma=2.0,1.0,0.5$ corresponding to strong, intermediate
  and weak coupling to the leads.}
\end{figure}

\subsection{Equilibrium spectral function}\label{sec.eqspec}
In Fig.~\ref{fig.spec} we show the $\varepsilon_c$-dependence of the
equilibrium spectral function, $A(\omega)=-\text{Im}G^r(\omega)$, for $U=4$ and $\Gamma=0.65$.
The HF solutions are Lorentzians centered at $\varepsilon_{\text{HF}}=\varepsilon_c+U\langle \hat
n_{\sigma} \rangle$ with a full width at half maximum (FWHM) given by
$2\Gamma$. As can be seen the position of the HF peaks do not vary
linearly with $\varepsilon_c$. Instead there is a "charging resistance" for the peak to move
through the Fermi level due to the cost in Hartree energy
associated with the filling of the level. This effectively pins
the level to $E_F$. 

Moving from HF to the second Born approximation, the Lorentzian shape
of the spectral peak is distorted due to the $\omega$-dependence of
the $2B$ self-energy. We can observe a general shift of spectral
weight towards the chemical potential as well as a narrowing of the resonance
as it comes closer to $E_F$.

The redistribution of the spectral weight towards the chemical potential becomes even more
pronounced in the $GW$ approximation. For $\Gamma-U<\varepsilon_c<-\Gamma$ (the so-called
Kondo regime) a sharp peak develops at $E_F$. For $U/\Gamma$ sufficiently
large the Kondo effect should reveal itself as a peak in the spectral function
with a full width at half maximum (FWHM) given approximately by the Kondo temperature~\cite{haldane} 
\begin{equation}\label{eq.tk}
T_K\approx 0.5(2\Gamma
U)^{1/2}\exp[\pi\varepsilon_c(\varepsilon_c+U)/2\Gamma U].
\end{equation} 
In Fig.~\ref{fig.tk} we compare the above expression for $T_K$ with the FWHM of the $GW$ Kondo peak. The exponential
scaling of $T_K$ is surprisingly well reproduced. Deviations from the
exponential scaling naturally occur for smaller values of $U/\Gamma$ (not shown)
where the Kondo effect does not occur and (\ref{eq.tk}) does not
apply. In accordance with recent
work~\cite{wang07}, we were not
able to obtain non-magnetic $GW$ solutions in the strong interaction
regime ($U/\Gamma > 8$).

In Fig.~\ref{fig.spec2} we show the dependence of the spectral
function on the ratio $U/\Gamma$ for the central level at the
symmetric position $\varepsilon_c=-U/2=-2$. For $U/\Gamma=2$
there is no significant difference between the three
descriptions. This is to be expected since the correlation plays a
minor role compared to the hybridization effects. In the weakly coupled
limit, however, correlations become significant and as a consequence the 2B and $GW$ results
changes markedly from the Lorentzian shape and show a Kondo-like
peak at the metal Fermi level. The 2B approximation significantly
overestimates the width of the Kondo peak, indicating, as expected,
that the higher order RPA terms enhance the strong correlation features.

For large $U/\Gamma$ it is known~\cite{wilson75,costi94} that the
spectral function, in addition to the Kondo peak, should develop peaks at the atomic
levels $\varepsilon_c$ and $\varepsilon_c+U$. We find that the self-consistent 2B and
$GW$ approximations always fail to capture these side-bands and instead distribute
the spectral weight as a broad slowly decaying tail. These findings
agree well with previous results obtained with the
fluctuation-exchange approximation~\cite{white92}, and with $GW$
studies of the homogeneous electron gas which showed that
self-consistency in the $GW$ self-energy washed out the satelite structure
in the spectrum~\cite{holm_barth}.

\subsection{Non-equilibrium transport}
\begin{figure}[!b]
\includegraphics[width=0.83\linewidth,angle=270]{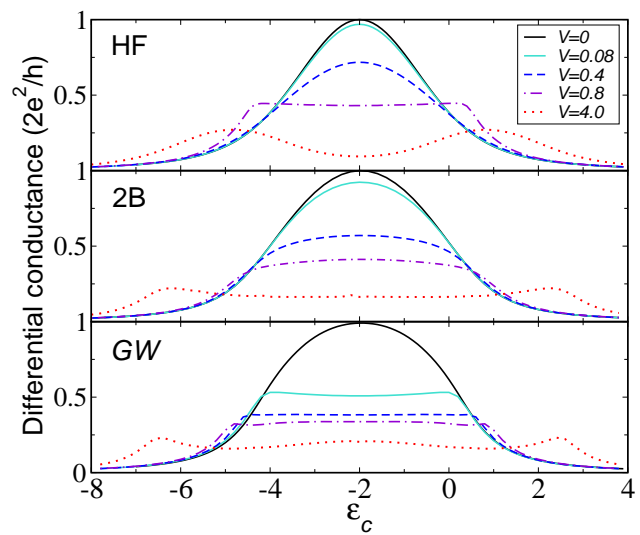}
\caption[cap.wavefct]{\label{fig.bias} (color online). Differential
  conductance, $\text dI/\text dV$, as a
  function of the central site energy, $\varepsilon_c$, for different
  applied biases, $U=4$ and $\Gamma=0.65$.}
\end{figure}

We now move to the non-equilibrium case and introduce a
difference in the chemical potentials of the two leads. In Fig.~\ref{fig.bias} we show the zero-temperature differential
conductance under a symmetric bias, $\mu_{L/R}=\pm V/2$, as a function
of $\varepsilon_c$ for $U=4$ and $\Gamma=0.65$. The $\text d I/\text d V$ at bias voltage $V$ has been calculated as a finite difference
between the currents obtained from Eq.~(\ref{eq.symcurrent}) for 
bias voltages $V$ and $V+\delta V$, respectively. The 2B result falls in between the HF and $GW$
results, and for this reason we will focus on the latter two in the
following discussion.

For $V=0$ there is only little difference between
the three results which all show a broad conductance peak
reaching the unitary limit at the symmetric point
$\varepsilon_c=-U/2$. The physical origin of the conductance trace is,
however, very different: While the HF result is
produced by coherent transport through a broad spectral peak
moving rigidly through the Fermi level, the $GW$ result is due to
transport through a narrow Kondo peak which is always on resonance
(for $\varepsilon_c$ in the Kondo regime). In all cases the width of the $\text d I/\text d V$
curve is approximately $U$. In the $GW$ case this is
because the Kondo peak develops only when the central level is half
occupied, i.e. $-U\lesssim \varepsilon_c \lesssim 0$. In HF, on the other hand, the
$\text d I/\text d V$ peak acquires a width on the order of $U$ due to the charge pinning
effect discussed in Sec.~\ref{sec.eqspec}.

The difference in the mechanisms leading to the HF and $GW$ results is brought
out clearly as $V$ is increased: for $V\ll \Gamma$ the bias has little
effect on the HF conductance while the $GW$ conductance drops
dramatically already at biases comparable to $T_K$ due to suppression
of the Kondo resonance at finite bias. The suppression of the Kondo
resonance is due to quasi-particle (QP) scattering. While QP
scattering does not
affect the life-time of QPs at $E_F$ in equilibrium, it does
so at finite bias where $\text{Im}\Sigma_{GW}(E_F)$ becomes
non-zero. We mention that we do not observe a
splitting of the $GW$ Kondo resonance at finite $V$~\cite{meir93}. 

The peaks appearing in the $\text d I/\text d V$ at the largest bias ($V=4$) occur when the
central level is aligned with either the lower or upper edge of the
bias window. It is worth noticing that the height of these peaks are
smaller than the value of $1G_0$ expected from on-resonant transport
through a single level. The reason for this is two-fold: (i) The
bias window only hits the resonance with one edge (either upper or
lower edge), and consequently only half the spectral
weight enters the bias window when the voltage is increased by $\Delta
V$ as compared to the low-bias situation. (ii) The self-consistent charging
resistance discussed in Sec.\ref{sec.eqspec} pins the level to
the edge of the bias window making the resonance follow the
bias. 

\subsection{The $G_0W_0$ approximation}\label{sec.nonself}
Non self-consistent, or one-shot, $GW$ calculations can be performed by
evaluating the screened interaction and $GW$ self-energy from some trial
non-interacting Green's function, $G_0$. The resulting $G_0W_0$
approximation, with $G_0$ obtained from an LDA/GGA calculation, has been found to yield very satisfactory results for
the band gaps of insulators and
semi-conductors~\cite{hybertsen86,rmp}. For this reason, and due to
its significantly lower computational cost, this $G_0W_0$ approach has generally been
preferred over the self-consistent $GW$. One rather unsatisfactory feature of the perturbative $G_0W_0$ method is its
$G_0$-dependence. However, as will be demonstrated below, a just as critical
problem in non-equilibrium situations is its 
non-conserving nature. 

Before we apply the $G_0W_0$ approximation to the Anderson model, we
need to address a certain issue which unfortunately has led to an
error in our previous paper Ref.~\onlinecite{thygesen_gw} (all
conclusions from that paper are, however, unaffected by the mistake.)

\begin{figure}[!h]
\includegraphics[width=0.8\linewidth,angle=270]{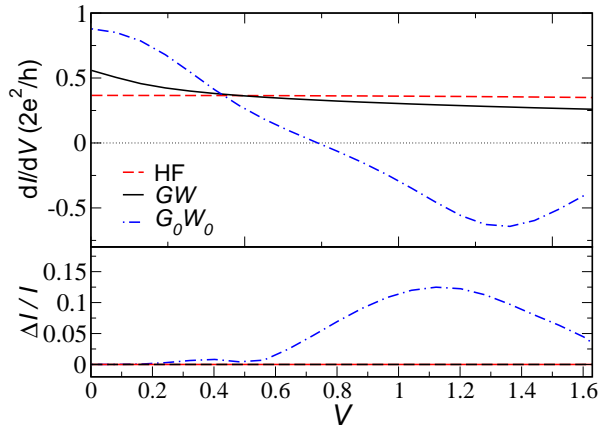}
\caption[cap.wavefct]{\label{fig.bias0} (color online). Differential
  conductance as a function of applied bias for $U=4$,
  $\Gamma=0.65$ and $\varepsilon_c=-4$. For these parameters, the
  non-magnetic HF solution is stable for bias voltages smaller than $\sim
  1.6$. The $G_0W_0$ approximation yields different
  currents at the left and right interfaces ($\Delta I\neq 0$), and
  yields negative differential conductance at finite bias.}
\end{figure}

\subsubsection{Instability of the non-magnetic ground state}
Consider a system which admits a spin polarized groundstate at the
Hartree level (notice
that Hartree and HF is equivalent for the Anderson model when the
effective interaction of Eq.~(\ref{eq.effint}) is used), and let $G_0$ denote the GF obtained from
spin unpolarized Hartree calculation. It turns out that the analytical
properties of the screenined interaction, $W^r_0[G_0]$, evaluated from
$G_0$ will be wrong. In particular $W^r_0[G_0]$ will not be retarded as
it should be. The reason is that the RPA response function is ill
defined around the non-magnetic, and thus unstable, $G_0$. The problem
has been previously mentioned by J. A. White~\cite{white92}, and was brought to the authors attention by
C. Spataru.

For certain parameter values, the HF groundstate of the Anderson model develops a finite magnetic
moment. As
a consequence the analytic
properties of $W_0^r$ as calculated from Eq.~(\ref{eq.wr}) with the
\emph{unpolarized} $G_{\text{HF}}$ become
wrong. In our previous
paper Ref.~\onlinecite{thygesen_gw}, this problem was not recognized because we,
for numerical efficiency, applied the Kramers-Kronig relation
(\ref{eq.corrr}) to obtain $\Sigma^r$ from $\Sigma^<-\Sigma^>$,
instead of using Eq.~(\ref{eq.sigmar}). Thus by construction our $\Sigma^r$
was retarded. Specifically, this implies that the $G_0W_0$ spectral
function plotted in Fig. 1 of that paper, as well as the
$\text{d}I/\text{d}V$ curves in the middle panel of Fig. 2 for
$\varepsilon_c$ in the
interval $-3.6$ to $-0.4$, are incorrect. In fact there exists no
non-magnetic $G_0W_0[G_{\text{HF}}]$ solution in these cases. We stress, however, that
all conclusions from our paper are unaffected by this mistake. In
particular we show below that for parameter
values leading to a \emph{stable} non-magnetic HF groundstate, the
$G_0W_0$ approximation
still violates charge conservation and gives unphysical results such as negative
differential conductance. Moreover, we arrive at the same conclusions
for $G_0W_0$ self-energies constructed from the
\emph{spin polarized} HF Green's function, in which case the
instability problem does not occur at all.

\subsubsection{Results of the $G_0W_0$ approximation}
In Fig.~\ref{fig.bias0} we show the calculated
$\text d I/\text d V$ for the Anderson model with $\Gamma=0.65$ and
$\varepsilon_c=-4$ for the HF, $GW$, and $G_0W_0[G_{\text{HF}}]$ approximations. For these parameters, the non-magnetic HF solution
is stable for bias voltages smaller than $\sim
  1.6$, such that the $G_0W_0$ approximation based on a non-magnetic
  $G_{\text{HF}}$ is indeed meaningful in this
  parameter range. The $G_0W_0$ conductance has been obtained as a
  finite difference between the currents obtained from Green's functions
  with self-energies $\Sigma_{GW}[G_{\text{HF}}(V)]$ and
  $\Sigma_{GW}[G_{\text{HF}}(V+\delta V)]$, respectively, where
  $G_{\text{HF}}(V)$ is the HF Green's function evaluated
  self-consistently under a bias voltage $V$.

\begin{figure}[!h]
\includegraphics[width=1.1\linewidth,angle=270]{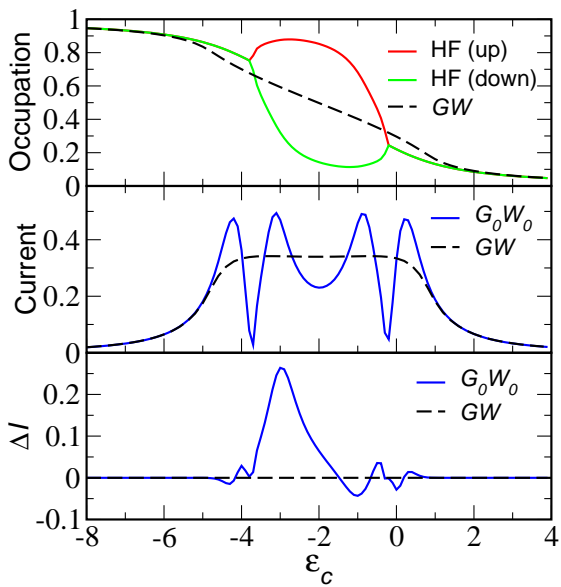}
\caption[cap.wavefct]{\label{fig.bias1} (color online). Upper panel:
  Occupation of the central site as function of $\varepsilon_c$ for $U=4$, $\Gamma=0.65$, and bias $V=0.8$. Notice that the HF solution breaks
  the spin symmetry for some $\varepsilon_c$ values. Middle panel:
  Current calculated in self-consistent $GW$ and
  $G_0W_0[G_{\text{HF},\uparrow},G_{\text{HF},\downarrow}]$. Lower
  panel: Violation of the continuity equation measured as the
  difference between the currents in the left and right leads.}
\end{figure}

\begin{figure}[!h]
\includegraphics[width=1.1\linewidth,angle=270]{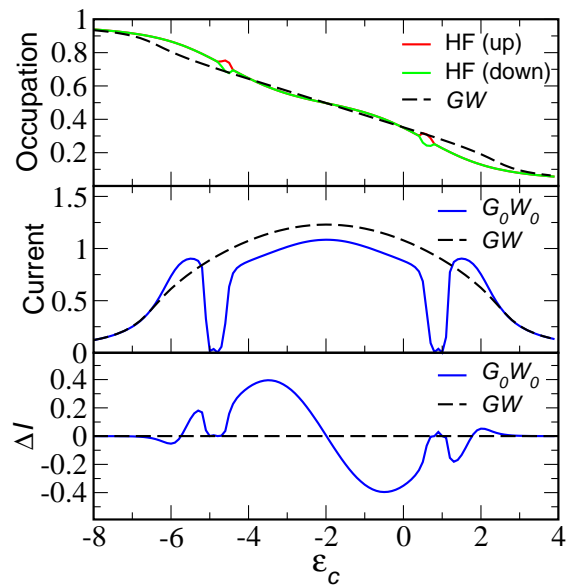}
\caption[cap.wavefct]{\label{fig.bias2} (color online). Same as
  Fig.~\ref{fig.bias1}, but for bias voltage $V=4.0$.}
\end{figure}

From Fig.~\ref{fig.bias0} we conclude that the
$G_0W_0$ approximation leads to
unphysical results in the form of strong negative differential
conductance. Moreover, as shown in the lower panel of the figure, the
$G_0W_0$ approach gives different values for $I_L$ and $I_R$. We note
in passing that this symmetry break comes from the different chemical
potentials of the left and right leads. Finally, we mention that the
increasing behavior of $\Delta I/I$ as function of bias voltage seems
to be a general effect.

As already mentioned the HF solution breaks the spin symmetry for
certain parameter values. Meaningfull $G_0W_0$ results can still be
obtained in this case provided the self-energy is constructed from the
spin polarized HF Green's function. Figs. \ref{fig.bias1} and
\ref{fig.bias2} compare the result of such calculations with
self-consistent $GW$ for two different
values of the bias voltage. From the figures we draw the conclusions:
(i) The $G_0W_0$ and $GW$ currents agree when the level is alomst
empty/filled (ii) The current calculated in $G_0W_0$ show
unphysical behavior in and close to magnetic regime (iii) The
violation of charge conservation in $G_0W_0$ is more severe when the
current is large.

\section{Benzene junction}\label{sec.benzene}
In this section we apply the Wannier-$GW$ method to a more realistic
nano junction, namely a benzene molecule 
coupled to featureless leads. In contrast
to the Anderson model considered in the preceding section, the benzene junction represents a closed-shell
system with the Fermi level lying within the HOMO-LUMO gap leading to rather
low transmission for all but the strongest molecule-lead coupling strengths.

The use of featureless (wide-band)
electrodes is convenient as it
allows us to isolate the effects of the electron-electron interactions.
The
use of more realistic contacts with energy dependent spectral features would lead to an additional renormalization of the
molecular levels making a clear separation between xc- and contact
effects more difficult. We stress, however, that the contacts only enter the
theory through the coupling self-energies which can be calculated once
and for all as in the standard NEGF-DFT approach. Thus the use of
more realistic contact self-energies is straightforward.

To describe the benzene molecule we first perform a DFT calculation
for the isolated molecule,
see note ~\onlinecite{details}. The KS eigenstates are then transformed
into maximally localized WFs, and the KS Hamiltonian and Coulomb
integrals are evaluated in the WF basis. For the interactions we use
the truncation scheme $\hat V^{(2)}$ defined in App.
\ref{app.hartree_exchange} to evaluate Hartree and exchange
self-energies. As shown in table~\ref{tab1} this leads to results
within $\sim 5\%$ of the exact
values. We use the effective interaction
Eq. (\ref{eq.effint}) for the correlation part of the $GW$ self-energy.
In all calculations we have applied a frequency grid extending from -100 to 100 eV,
and grid spacings in the range 0.2 to 0.02, depending on the value of $\Gamma$. 

In Sec.~\ref{sec.spectrum} we show that the experimental ionization potential of
the isolated benzene molecule is very well reproduced with our $GW$
scheme. In Sec.~\ref{sec.gamma} we investigate the role of the
coupling strength, $\Gamma$, on the spectrum of the benzene
junction. Finally, in
Sec.~\ref{sec.benzene_noneq} we calculate the non-equilibrium
conductance of the junction and compare various approximations for the
xc self-energy.

\begin{figure}[!b]
\includegraphics[width=1.0\linewidth]{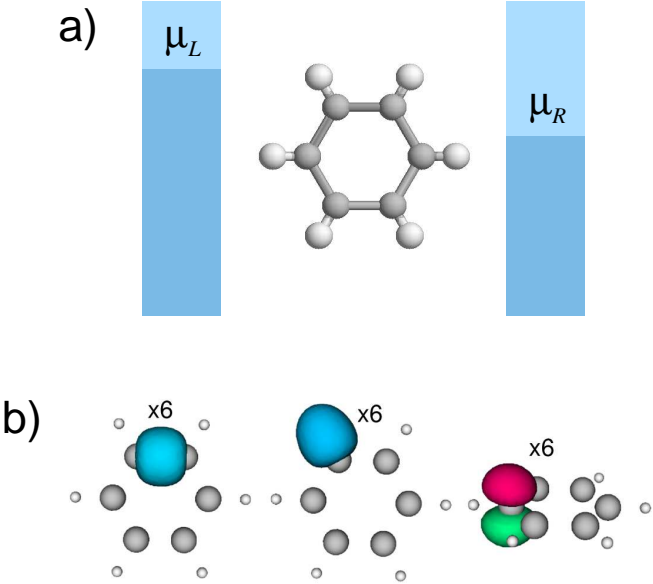}
\caption[cap.wavefct]{\label{fig.benzene} (color online). (a)
  Illustration of a benzene molecule coupled to featureless electrodes
  with different chemical potentials. (b) Iso-surfaces
  for the 18 partially occupied Wannier functions used as basis
  functions in the calculations. The WFs are linear combinations of 
  Kohn-Sham eigenstates obtained from a DFT-PBE plane-wave calculation. 
}
\end{figure}

\subsection{Spectrum of isolated benzene}\label{sec.spectrum}
Within our general transport formalism we model the situation of a
free molecule by using a very weak coupling to the wide-band leads,
see Fig.\ref{fig.benzene}(a). The contacts merely act as particle reservoirs fixing the number of electrons on the
molecule and providing an insignificant broadening ($\Gamma=0.05$eV) of the discrete
energy levels. We fix the Fermi
levels of the electrodes to $E_F=-3$~eV which is approximately
half-way between the HOMO and LUMO levels (the precise position of
$E_F$ within the gap is unimportant for the results presented in this
section). 

\begin{figure}[!b]
\includegraphics[width=0.8\linewidth,angle=270]{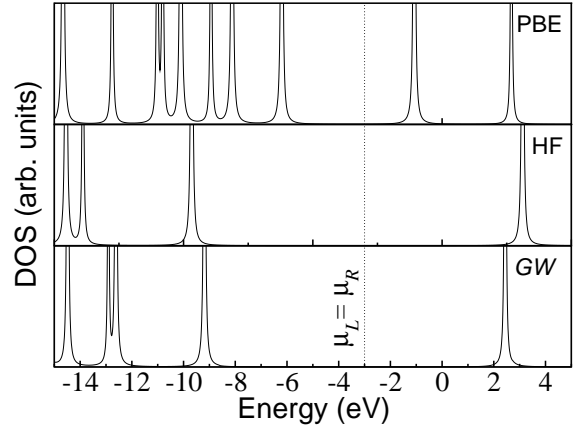}
\caption[cap.wavefct]{\label{fig.c6h6_spec} Density of states for a benzene molecule weakly coupled to
  featureless leads ($\Gamma=0.05$). The common Fermi levels of the leads is
  indicated. Notice the characteristic opening of the band gap 
  when going from DFT-PBE to HF, and the subsequent (slight) reduction when
  correlations are included at the $GW$ level.}
\end{figure}

\begin{figure}[!b]
\includegraphics[width=0.8\linewidth,angle=270]{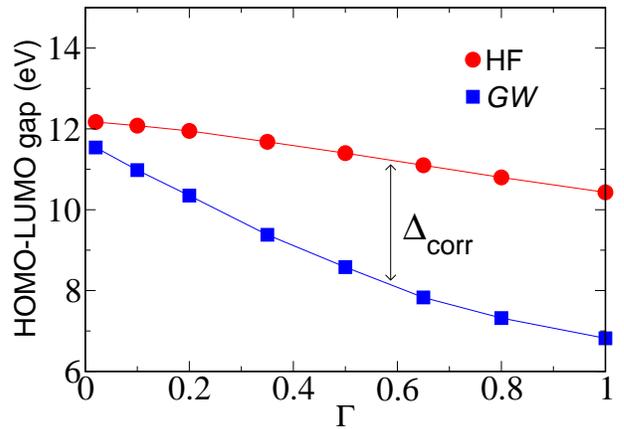}
\caption[cap.wavefct]{\label{fig.gap_vs_gamma} The HF and $GW$ HOMO-LUMO gap of
  the benzene molecule as a function of the coupling strength
  $\Gamma$. The difference between the curves represents the reduction
in the gap due to the correlation part of the $GW$ self-energy. This
value increases with the coupling strength, as screening by electrons
in the leads becomes more effective.}
\end{figure}

In Fig.~\ref{fig.c6h6_spec} we show the total density of states (DOS),
\begin{equation}
D(\varepsilon)=-\frac{1}{\pi}\sum_{n=1}^{N_w} \text{Im}G^r_{nn}(\varepsilon) 
\end{equation},
where the sum runs over all WFs on the molecule. We use three
different approximations: (i) DFT-PBE (ii)
Hartree-Fock (iii) fully self-consistent $GW$. We stress that our
calculations include the full dynamical dependence of the $GW$
self-energy as well as all off-diagonal
elements. Thus no analytic extension is performed, and we do not
linearize the self-energy around
the DFT eigenvalues to obtain an approximate quasi-particle equation
as is done in standard $GW$ calculations. 

The spectral peaks seen in Fig.~\ref{fig.c6h6_spec} occurring above
(below) the Fermi level correspond to electron addition (removal)
energies. In particular, the HOMO level should coincide with the
(vertical) ionization energy of the isolated molecule, which in the
case of benzene is $I_{\text{exp}}=-9.2$~eV~\cite{NIST}. The PBE
functional overestimates this value by 3~eV, giving
$I_{\text{PBE}}=-6.2$~eV in good agreement with previous
calculations~\cite{niehaus}. The HF and $GW$ calculation yields
$I_{\text{HF}}=-9.7$~eV and $I_{GW}=-9.3$~eV, respectively. Because of
the limited size of the WF basis, the perfect agreement between the
$GW$ and experimental values should not be taken too strict. Indeed,
more accurate HF calculations predicts a HOMO level around $-9.2$~eV
which is $0.5$~eV higher than our HF result.
The deviation of our HF calculation from this number is two-fold: (i)
The use of the truncated interaction $V^{(2)}$ to evaluate the
exchange self-energy introduces an error of $\sim 0.1$~eV, see
table~\ref{tab1}. (ii) The difference between the PBE
orbitals (from which our WFs are constructed) and the true HF
orbitals. 

Returning to Fig.~\ref{fig.c6h6_spec},
we notice a dramatic opening of the HOMO-LUMO gap when going from PBE
to HF (and $GW$). This effect is due to the inability of the LDA/GGA functionals
to fully cancel the spurious self-interaction contained in the Hartree
potential. For the same reason, the self-interaction free HF method
generally yields better spectra than the LDA/GGA functionals for small, localized systems where 
self-interaction terms are significant and dynamic screening is small.
The $GW$ spectrum resembles the HF spectrum with a slight reduction of
the gap by $\sim 1.0$~eV. As we show in the next section, the $GW$ gap 
shrinks as the coupling strength, $\Gamma$, is increased.

\subsection{Contact enhanced screening (the role of $\Gamma$)}\label{sec.gamma}
In Fig. \ref{fig.gap_vs_gamma} we plot the size of the HOMO-LUMO gap
as a function of the coupling strength $\Gamma$. Both the HF and $GW$
gaps decrease as $\Gamma$ is increased. For the HF gap, this is a simple 
consequence of the redistribution of charge from the HOMO to the LUMO when the
resonances broaden and their tails start to cross the Fermi
level. As this happens the HOMO (LUMO) self-interaction term in
$\Sigma_x$ will become less (more) negative, and consequently the HF gap shrinks.

The $GW$ quasi-particle energies consist of a HF eigenvalue and a
correlation contribution coming from the real part of the dynamic $GW$ self-energy, 
\begin{equation}
\varepsilon^n_{\text{QP}}=\varepsilon^n_{\text{HF}}+\Delta_{\text{corr}}^n.
\end{equation}
According to Fig.~\ref{fig.gap_vs_gamma}, the correlation part of the
QP gap,
\begin{equation}
\Delta_{\text{corr}}=\Delta_{\text{corr}}^{\text{HOMO}}-\Delta_{\text{corr}}^{\text{LUMO}},
\end{equation}
increases significantly with $\Gamma$. In fact for a large range of
coupling strengths, the reduction of the gap is more than 3~eV.  This
reduction can be understood from the enhanced mobility of the
electrons on the molecule when the coupling is strong. The enhanced
mobility allows for more efficient screening and this reduces the QP
gap. The difference between the large- and small $\Gamma$ limits
is analogue to the difference between extended and confined
systems. In extended systems where screening is significant, band gaps are
overestimated by HF, and correlation contributions to the gap are
large. In confined systems, such as atoms and small molecules, screening effects
are unimportant and HF usually yields good HOMO-LUMO gaps.

\begin{figure}[!h]
\includegraphics[width=0.85\linewidth,angle=270]{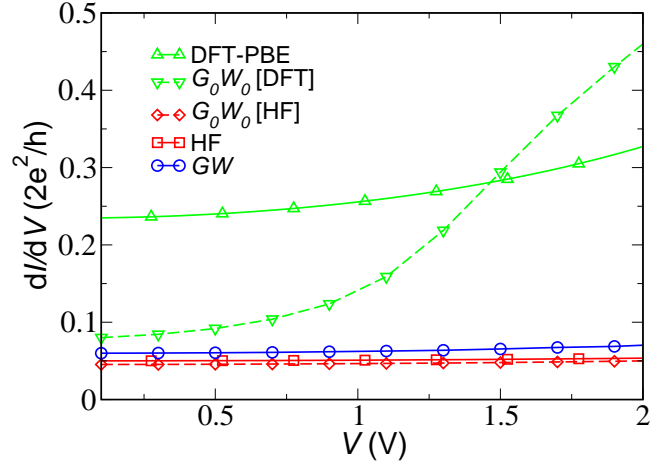}
\caption[cap.wavefct]{\label{fig.benzene_cond} (color
  online). Differential conductance of the benzene junction for
  $\Gamma_L=\Gamma_R=0.25$~eV. Notice the strong $G_0$ dependence of the
  $G_0W_0$ result.}
\end{figure}

\begin{figure}[!h]
\includegraphics[width=0.85\linewidth,angle=270]{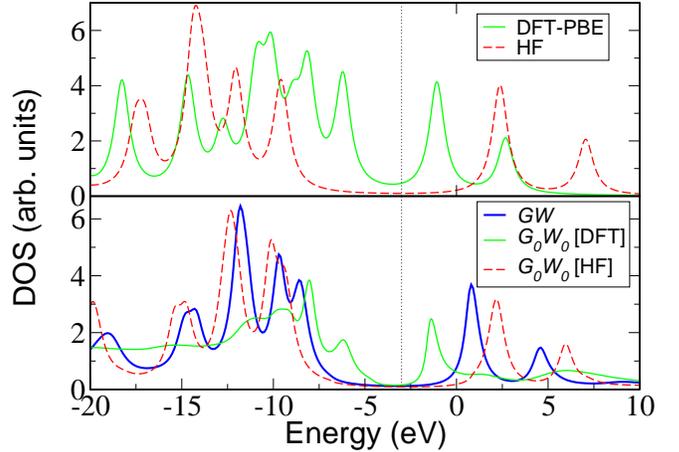}
\caption[cap.wavefct]{\label{fig.benzene_dos} (color online). Equilibrium DOS
  for the benzene molecule coupled to wide-band leads with a coupling strength of
  $\Gamma_L=\Gamma_R=0.25$~eV. Upper panel shows DFT-PBE and HF single-particle
  approximations while the lower panel shows the self-consistent $GW$
  result as well as one-shot $G_0W_0$ results based on the DFT and HF
  Green's functions, respectively.}
\end{figure}

\subsection{Conductance}\label{sec.benzene_noneq}
In this section we consider the transport properties of the benzene junction under
a symmetric bias, $\mu_{L/R}=\pm V/2$, and a coupling strength of
$\Gamma_L=\Gamma_R=0.25$~eV. 

In Fig.~\ref{fig.benzene_cond} we compare the differential conductance,
$\text dI/\text d V$, calculated in self-consistent DFT-PBE, HF, $GW$, as well as
non self-consistent $G_0W_0$ using either the DFT-PBE or HF Green's
function as $G_0$. The $\text dI/\text dV$ has been obtained by numerical differentiation of the $I(V)$ curves
calculated from Eq.~(\ref{eq.symcurrent}). For the  DFT calculation
the finite-bias effects have been included at the Hartree level,
i.e. changes in the xc-potential have been neglected. 
We notice that the HF and $G_0W_0[G_{\text{HF}}]$ results are close to
the self-consistent $GW$ result. These approximations all yield a
nearly linear $IV$ with a conductance of $\sim0.05G_0$. In contrast the DFT
and $G_0W_0[G_{\text{DFT}}]$ yield significantly larger conductances
which increase with the bias voltage. We note that the violation of charge
conservation in the $G_0W_0$ calculations is not too large in the
present case ($\Delta I/I<5\%$). This is in line with our general
observation, e.g. from the Anderson model, that $\Delta I/I$ grows with $I$.

The trends in conductance can be understood by considering the
(equilibrium) DOS of the junction shown in
Fig.~\ref{fig.benzene_dos}. As for the free benzene molecule (see
Fig.~\ref{fig.c6h6_spec}) the
DFT HOMO-LUMO gap is much smaller than the HF gap, and this explains
the lower conductance found in the latter case. The $GW$ gap falls in
between the DFT and HF gaps, however, the magnitude of the DOS at
$E_F$ is very similar in $GW$ and HF which is the reason for the
similar conductances. It is 
interesting to notice that the HOMO-LUMO gap obtained in the $G_0W_0$
calculations resemble the gap obtained from $G_0$, and that the
self-consistent $GW$ gap lies in between the $G_0W_0[G_{\text{DFT}}]$ and
$G_0W_0[G_{\text{HF}}]$ gaps.

The increase in the $G_0W_0[G_{\text{DFT}}]$ conductance as a function of
bias occurs because the LUMO of the $G_0W_0[G_{\text{DFT}}]$ calculation
moves downwards into the bias window and becomes partly filled as the voltage is raised. In a
self-consistent calculation this would lead to an increase in Hartree
potential which would in turn raise the energy of the level. The
latter effect is missing in the perturbative $G_0W_0$ approach and this
can lead to uncontrolled changes in the occupations as the present
example shows.

Finally, we notice that the $G_0W_0[G_{\text{DFT}}]$ DOS is significantly more
broadened than both the $G_0W_0[G_{\text{HF}}]$ and $GW$ DOS. The reason
for this is that the DFT DOS has a relatively large weight close to
$E_F$. This enhances the QP scattering and leads to shorter life-times
of the QP in the $G_0W_0[G_{\text{DFT}}]$ calculation. Noticing that
the QP life-time is inversely proportional to
$\text{Im}\Sigma_{GW}$ this explains the broadening of the spectrum.

\section{Conclusions}\label{sec.conclusions}
With the aim of investigating the role of electronic correlations in quantum
transport, we have implemented the non-equilibrium $GW$ approximation
to the electronic self-energy of a finite region of interacting
electrons coupled to non-interacting leads. We have shown, both analytically and by means of numerical
examples, that the self-consistent $GW$ self-energy leads to identical
currents at the left and right interfaces of the central region.
In contrast, the widely used $G_0W_0$ self-energy does not conserve particle number and thus violates
the continuity equation. More generally we have shown that any
$\Phi$-derivable self-energy will yield identical left- and right-
currents independent of the basis set applied. 

Using a WF basis we have introduced an effective electron-electron interaction which resembles the real space
representation but is spin-dependent and self-interaction free in the
WF basis. In general this provides a means for reducing
self-interaction errors in diagrammatic approaches like the $GW$
method.

The $GW$ method was applied to the Anderson impurity model. In
equilibrium and $T=0$ we found that the self-consistent $GW$ approximation
describes the width of the Kondo resonance well for intermediate
interaction strengths, $U=4$ and $\Gamma\gtrsim 0.5$. On the other
hand the sidebands of the spectral
function are always missed in $GW$. We presented non-equilibrium $IV$ curves
and discussed the important effect of quasi-particle scattering under
finite bias which reduce the QP life-times leading to a broadening of
spectral features and significant suppression of the finite bias conductance. Finally, we demonstrated that
the $G_0W_0$ approach can produce severe errors
including violation of charge conservation and negative differential
conductance. The errors become more significant at higher bias and close to magnetic transition points.

We investigated the properties of a molecular junction
consisting of a benzene molecule sandwiched between featureless leads.
To describe the benzene we used a minimal Wannier function basis set
which was shown to reproduce the exact Hartree and exchange matrix
elements to within $5\%$. The calculated ionization potential in $GW$ 
was found to be in good agreement with the experimental value.
A significant reduction of the $GW$ HOMO-LUMO gap was observed for
increasing molecule-lead coupling. The effect comes from the
correlation part of the $GW$ self-energy and reflects the more
efficient screening in a strongly, compared to a weakly, coupled junction.

Finally, the non-equilibrium differential conductance of the benzene
junction was calculated in DFT-PBE, HF, and $GW$ as well as
$G_0W_0[\text{HF}]$ and $G_0W_0[\text{DFT}]$. It was found that HF,
and $G_0W_0[\text{HF}]$ yield results similar to $GW$, while both DFT
and $G_0W_0[\text{DFT}]$ yield significantly larger conductances. In
particular, this shows that the $G_0$-dependence of the $G_0W_0$
approximation should not be disregarded. The trends in conductance
were explained in terms of the size of the HOMO-LUMO gap of the
molecule which also shows significant variation depending on the
approximation used.

\section{Acknowledgments}
The authors thank E.K.U Gross, S. Kurth, G.
Stefanucci, R. Godby, A. Feretti, A. -P. Jauho, and K. Kaasbjerg for
useful discussions. We thank Catalin Spataru for pointing out the problem related
to the instability of non-magnetic HF groundstate, and Carsten
Rostgaard for discussions concerning the evaluation of Coulomb
integrals. K. S. Thygesen acknowledge support from the Danish
Natural Science Research Council, and from the Danish Center for
Scientific Computing through grant No. HDW-1103-06. The Center for Atomic-scale
Materials Design (CAMD) is sponsored by the Lundbeck Foundation.
A. Rubio acknowledges support from the EC Network of Excellence NANOQUANTA
(ref. NMP4-CT-2004-500198), the Spanish Ministry of Education (grant
FIS2007-65702-C02-01), the SANES (ref. NMP4-CT-2006-017310),
DNA-NANODEVICES (ref. IST-2006-029192), and NANO-ERA Chemistry projects,
the University of the Basque Country EHU/UPV (SGIker Arina),  Basque
Country Government and  the
computer resources, technical expertise and assistance provided by the
Barcelona Supercomputing Center - Centro Nacional de Supercomputaci\'{o}n.

\appendix
\section{Hartree and exchange potentials}\label{app.hartree_exchange}
In this work the exchange and Hartree self-energies have been
evaluated from Eqs.~(\ref{eq.hr}) and (\ref{eq.exr}) with the Coulomb
matrix elements restricted to a certain subset (the set $\hat V^{(2)}$ defined
below). Here we
investigate the quality of such approximations by testing their
ability to reproduce Hartree and exchange energies of the
molecular orbitals of a benzene molecule~\cite{details}. We thus consider the
following truncation schemes 
\begin{eqnarray}\label{eq.intapprox2}
\hat V^{(1)} &=& \hat V[\:\{V_{ij,ij}\},\{V_{ij,ji}\}\:],\\ \label{eq.intapprox3}
\hat V^{(2)} &=& \hat
V[\:\{V_{ij,ij}\},\{V_{ij,ji}\},\{V_{ii,jj}\},\{V_{ii,ij}\}\:],\\
\label{eq.intapprox4}
\hat V^{(3)} &=& \hat V[\:\{V_{ij,ij}\},\{V_{ij,ji}\},\{V_{ii,jj}\},\{V_{ii,ij}\},\{V_{ik,jk}\}\:],
\end{eqnarray}
where e.g. the notation $\hat V[\{V_{ij,ij}\}]$ means that all elements of
the form $V_{ij,ij}$ are included in the sum in Eq.
(\ref{eq.interaction}). 

The
molecular orbitals of benzene, $\{\psi_n\}$, can, by construction of the
WFs $\{\phi_i\}$, be exactly expanded as,
\begin{equation}\label{eq.psinexpan}
\psi_n(\bold r)=\sum_i c_{in}\phi_i(\bold r).
\end{equation}
The 18 WFs used to describe the benzene molecule are plotted in Fig.~\ref{fig.benzene}(b).
For the molecular orbital $\psi_n$ we can then calculate the exact Hartree and exchange energies from
\begin{eqnarray}\nonumber
\langle \psi_n|\Sigma_h|\psi_n\rangle&=&2\sum_m^{\text{occ}}\int \text{d}\bold r \text{d}\bold r' \frac{\psi_n(\bold r)^*\psi_m(\bold r')^*\psi_m(\bold r')\psi_n(\bold r)}{|\bold r-\bold r'|}\\ \nonumber
\langle \psi_n|\Sigma_x|\psi_n\rangle&=&-\sum_m^{\text{occ}}\int \text{d}\bold r \text{d}\bold r' \frac{\psi_n(\bold r)^*\psi_m(\bold r')^*\psi_m(\bold r)\psi_n(\bold r')}{|\bold r-\bold r'|}
\end{eqnarray}
Alternatively we can insert the expansion (\ref{eq.psinexpan}) and get
\begin{eqnarray}
\langle \psi_n|\Sigma_h|\psi_n\rangle&=&\sum_{ij}c^*_{in}\Sigma_{h,ij}c_{jn}\\
\langle \psi_n|\Sigma_x|\psi_n\rangle&=&\sum_{ij}c^*_{in}\Sigma_{x,ij}c_{jn}
\end{eqnarray}
where $\Sigma_{h,ij}$ and $\Sigma_{x,ij}$ are the self-energies in
the WF basis obtained from
Eqs.~(\ref{eq.hr}),(\ref{eq.exr}). The latter are approximated by
the truncation schemes
(\ref{eq.intapprox2}-\ref{eq.intapprox4}) for the Coulomb
integrals, $V_{ij,kl}$. 

In table \ref{tab1} we
compare the exact values of the Hartree and exchange matrix elements
for the frontier molecular orbitals to the approximate ones obtained using the
truncated interactions. We note that $\hat V^{(2)}$, which is the
truncation scheme we have used, leads to average deviations around $5\%$. 

As a final remark we notice that our results for $\langle\psi_n
|\Sigma_x|\psi_n\rangle$ evaluated using $\hat V^{(1)}$ provides
roughly the same accuracy as a recently developed method combining
tight-binding DFT with $GW$\cite{niehaus}.

\begin{widetext}
\begin{center}
\begin{table}[h]
\begin{center}
\begin{tabular}{lccccccccccccc}
\hline
\hline
    &\vline & $\varepsilon_{\text{DFT}}$   &\vline & & &  $\langle\psi_n
    |\Sigma_h|\psi_n\rangle$ & &\vline & & $\langle\psi_n
    |\Sigma_x|\psi_n\rangle$ & &\\
         \hline                                       
State (symmetry) \quad  &\vline & &\vline&\quad  $\hat V^{(1)}$ \quad  &
\quad  $\hat V^{(2)}$ \quad & \quad  $\hat V^{(3)}$
\quad & \quad  Exact \quad  &\vline & \quad
$\hat V^{(1)}\quad $ &\quad  $\hat V^{(2)}$ \quad &\quad  $\hat V^{(3)}$\quad  &\quad Exact & \\
\hline
\hline
HOMO-2 ($\pi$)&\vline &-8.94 &\vline & 217.6 & 221.2 & 233.2 &233.0 &\vline&-14.6 & -16.4 & -16.3 &-16.7 &\\
\hline
HOMO-1 ($\sigma$)&\vline &-8.12 &\vline & 253.6 & 244.7 & 224.7 &224.6 &\vline&-24.7 & -20.2 & -20.2 &-19.6 &\\
\hline
HOMO ($\pi$)&\vline &-6.20 &\vline &  220.1 & 223.0 & 229.5 & 229.5 &\vline&-13.7 & -15.2 & -15.0 &-15.1  &\\
\hline
\hline
LUMO ($\pi^*$)&\vline &-1.08 &\vline & 222.1 & 222.7 & 219.1 & 219.3 &\vline&-7.2 & -7.5 & -7.5 &-7.2  &\\
\hline 
LUMO+1 ($\pi^*$)&\vline &2.68 &\vline & 223.1 & 221.4 & 199.3&199.7&\vline& -6.1 & -5.3 & -5.8 &-4.7  &\\
\hline
\hline
Average deviation ($\%$)& & &\vline & 7.3 & 5.8 &
0.1 & - &\vline& 15.5 & 4.5 & 6.7 & -  &\\
\hline
\hline
\end{tabular}
\end{center}
\caption{Hartree and exchange energies in eV for five frontier molecular
  orbitals of the benzene molecule. The values are obtained using the
  truncated interactions defined in
  Eqs.~(\ref{eq.intapprox2}-\ref{eq.intapprox4}) as well as the full interaction $\hat V$ (the exact result). For reference the first column shows the eigenvalues as calculated using the PBE xc-functional.}\label{tab1}
\end{table}
\end{center}
\end{widetext}

\section{Assessment of effective interaction}\label{app.effint}
As discussed in Sec.~\ref{sec.effint}, the $GW$ approximation includes only
a single diagram at each order of the interaction. The
error resulting from such an approximation is - to lowest order - similar to the error of approximating
a HF calculation by a Hartree calculation. It is not obvious that the
best result of such an approximation is obtained by using the full
interaction of Eq.~(\ref{eq.interaction}). For example such a
strategy would lead to self-interaction errors.

In table \ref{tab2} (middle panel) we compare Hartree matrix elements of some
molecular orbitals of benzene~\cite{details}, evaluated using different effective
interactions. Notice that the values listed in the two leftmost
columns differ by the inclusion of the spin dependent term of
Eq.~(\ref{eq.effint}). In the right column we show the exact HF
result, i.e. the correct result to first order in the interaction. The
last row shows the average deviation of the Hartree energies from the
exact HF energies.

From table \ref{tab2} we conclude that the effective interaction
produces results of comparable accuracy to the full interaction, if
one attempts to reproduce the exact result to first order from the
Hartree approximation only. The fact that $\hat V_{\text{eff}}$
performs better than $\{V_{ij,ij}\}$ indicate that the spin-dependent
term in $\hat V_{\text{eff}}$, which removes the self-interaction in
the WF basis, is significant.

Extrapolating these observations to higher order we
conclude that the use of $\hat
V_{\text{eff}}$ in $GW$ calculations should produce results comparable
to $GW$ calculations based on the full interaction.

At this point we stress again, that for practical calculations we use the truncation scheme of
Eq.~(\ref{eq.intapprox3}) for evaluating Hartree and exchange. Thus 
the results presented in this section only serve to estimate the performance of the
effective interaction for the higher-order $GW$ diagrams.

\begin{widetext}
\begin{center}
\begin{table}[h]
\begin{center}
\begin{tabular}{lcccccccc}
\hline
\hline
  &  \vline & & $\langle\psi_n
    |\Sigma_h|\psi_n\rangle$ & &\vline & $\langle\psi_n
    |\Sigma_h+\Sigma_x|\psi_n\rangle$ &\\
         \hline                                       
State (symmetry)   &\vline& $\{V_{ij,ij}\}$  &
  $\hat V_{\text{eff}}$ & Exact &\vline & Exact & \\
\hline
\hline
HOMO-2 ($\pi$)& \vline & 217.6  & 207.4 & 233.0 &\vline& 216.3 &\\
\hline
HOMO-1 ($\sigma$)&\vline & 253.9 & 230.1 & 224.6 & \vline& 205.0 &\\
\hline
HOMO ($\pi$)&\vline &  220.7 & 210.1 & 229.5 & \vline& 214.4  &\\
\hline
\hline
LUMO ($\pi^*$)&\vline & 222.8 & 212.2 &219.3 & \vline& 212.1  &\\
\hline 
LUMO+1 ($\pi^*$)&\vline & 223.7 & 213.1 & 199.7 &\vline& 195.0  &\\
\hline
\hline
Average deviation ($\%$) from exact HF &\vline & 9.5 & 5.5 & 6.0 & \vline& -  &\\
(right column) &\vline &  &  &  & \vline&   &\\
\hline
\hline
\end{tabular}
\end{center}
\caption{Left part: Hartree self-energy for
  some of the frontier orbitals of the benzene molecule. The Hartree
  self-energy has been evaluated using the effective interaction
  Eq.~(\ref{eq.intapprox}), the effective interaction without the
spin dependent correction (second term in
Eq.~(\ref{eq.effint})), and using the full interaction
Eq.~(\ref{eq.interaction}) (exact result). Right: The exact value of the Hartree-Fock
self-energy. Note that the spin dependent correction term in $\hat
V_{\text{eff}}$ cancels the self-interaction (in the local Wannier
basis) and thus incorporates part of the exchange in the
Hartree potential. Last row shows the average deviation of the Hartree potential from the exact Hartree-Fock potential.}\label{tab2}
\end{table}
\end{center}
\end{widetext}

\section{A useful relation}\label{app.useful}
Let $B(\tau,\tau')$ and $C(\tau,\tau')$ be two matrix valued functions
on the Keldysh contour, and consider the commutator $A$ defined by
\begin{equation}\label{app.defA}
A(\tau,\tau')=\int_{\mathcal C}[B(\tau,\tau_1)C(\tau_1,\tau')-C(\tau,\tau_1)B(\tau_1,\tau')]\text{d}\tau_1,
\end{equation} 
where matrix multiplication is implied. Under steady state conditions where the real time components of $B$ and $C$ can be assumed to depend only on the time difference $t'-t$, the following identity holds:
\begin{equation}\label{app.result}
\text{Tr}[A^<(t,t)]=\int \frac{\text{d}\omega}{2\pi} \text{Tr}[B^<(\omega)C^>(\omega)-B^>(\omega)C^<(\omega)].
\end{equation}
To prove this relation we first use the Langreth rules to obtain 
\begin{eqnarray}\nonumber
A^<(t,t')&=&\int \big [B^<(t,t_1) C^a(t_1,t') +B^r(t,t_1) C^<(t_1,t')\\ \nonumber &-&C^<(t,t_1) B^a(t_1,t') -C^r(t,t_1) B^<(t_1,t')\big ]\text{d}t_1.
\end{eqnarray}
Since all quantities on the right hand side depend only on the time difference we identify the integrals as convolutions which in turn become products when Fourier transformed. We thus have
\begin{eqnarray}\nonumber
A^<(t,t)&=&\int \frac{\text{d}\omega}{2\pi} A^<(\omega)\\ \nonumber
&=&\int \frac{\text{d}\omega}{2\pi} [B^<(\omega) C^a(\omega) +B^r(\omega) C^<(\omega)\\ \nonumber&&\quad -C^<(\omega) B^a(\omega) -C^r(\omega) B^<(\omega)].
\end{eqnarray} 
Eq.~(\ref{app.result}) now follows from the cyclic property of the trace and the identity $G^r-G^a=G^>-G^<$.

\section{Coupling to quasi-periodic leads}\label{sec.aperleads}
We consider the coupling of the central region ($C$) to the left lead
($L$) in the case where $L$ is periodic only beyond a certain
transition region ($T$). We refer to the periodic
parts of the lead as principal layers and denote the corresponding
blocks of the Hamiltonian matrix by $h_{0}$. Without loss of
generality we assume nearest
neighbor coupling between the principal layers and denote the coupling
matrices by $v_0$. The transition region is assumed so large that
there is no coupling across it, i.e. between the central region and
the first principal layer. If this is not the case the transition
region must be extended by the first principal layer. The Hamiltonian
of the left lead and its periodic part can then be written as
\begin{equation}
h_L=\left(\begin{array}{cccc}
        \ddots & \vdots & \vdots & \vdots  \\
        \hdots & h_0 & v_0 & 0 \\
        \hdots & v_{0}^{\dagger} & h_0 & v_{T} \\
        \hdots & 0 & v_{T}^{\dagger}& h_T \\
        \end{array}
\right),
\:
h_L^{per}=\left(\begin{array}{cccc}
        \ddots & \vdots & \vdots & \vdots  \\
        \hdots & h_0 & v_{0} & 0 \\
        \hdots & v_{0}^{\dagger} & h_0 & v_{0} \\
        \hdots & 0 & v_{0}^{\dagger}& h_0 \\
        \end{array}
\right)
\end{equation}
The (retarded) GFs defined from $h_L$ and $h_L^{per}$ are denoted by $g_{0,L}$ and $g^{per}_{0,L}$, respectively. 
The lower right block of $g_{0,L}$, corresponding to the transition region, is denoted $[g_{0,L}]_T$, and the lower right block of $g^{per}_{0,L}$, corresponding to the first principal layer, is denoted by $[g^{per}_{0,L}]_0$. We have the following equation
\begin{equation}
[g_{0,L}]_T=[(\omega+i\eta)I-h_{T}-\Sigma_{T}]^{-1}
\end{equation}
where the self-energy is given by
\begin{equation}
\Sigma_T=v_{T}^{\dagger}[g^{per}_{0,L}]_0 v_{T}.
\end{equation}
In the above equation $[g^{per}_{0,L}]_0$ can be obtained using the
standard decimation technique~\cite{guinea}. The coupling self-energy
$\Sigma_L$ can now be constructed from $[g_{0,L}]_T$ and the matrices
$h_{TC}$ and $h_{CT}$ which describe the coupling between the
transition region in the left lead and the central region,
\begin{equation}
\Sigma^r_L=h_{CT}[g^r_{0,L}]_T h_{TC}.
\end{equation} 
We remark that $h_{CT}$ and $h_{TC}$ are sub-matrices of $h_{CL}$ and $h_{LC}$. Completely analogue results hold for the coupling to the right lead.

\begin{figure}[!h]
\includegraphics[width=1.0\linewidth]{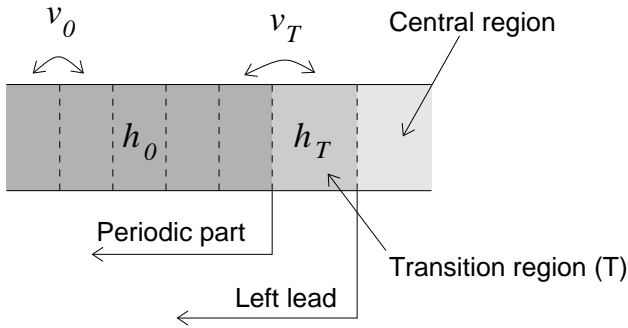}
\caption[cap.wavefct]{\label{fig.trans} The transition region, $T$, is defined as the part of the lead beyond which the lead Hamiltonian becomes periodic.}
\end{figure}

\section{Retarded functions from correlation functions}\label{sec.retfromcorr}
In steady-state all four real time GFs Eqs.(\ref{eq.gl})-(\ref{eq.adv})
follow from the retarded and lesser components and thus it suffices to
calculate these. 

Given $G^{r}(\omega)$ and $G^{<}(\omega)$ sampled on an equidistant frequency
grid, the corresponding $GW$ self-energy, $\Sigma_{GW}[G](\omega)$,
can be obtained from Eqs.~(\ref{eq.sigmar})-(\ref{eq.pl}) using the fast Fourier transform (FFT) to switch between energy and time
domains. However, as alternative to Eqs.~(\ref{eq.sigmar}) and (\ref{eq.pr}) we
have found it more useful to obtain $\Sigma_{GW}^r$ and $P^r$ from the relation 
\begin{equation}\label{eq.xrt}
X^r(t)=\theta(-t)[X^>(t)-X^<(t)],
\end{equation} 
which is valid for any function $X$ on the Keldysh contour that does not
contain delta functions. Note, that when applied to $\Sigma_{GW}$
Eq.~(\ref{eq.xrt}) yields only the correlation part of 
$\Sigma_{GW}^r$ as explained in Sec.~\ref{sec.noneqgw}.
The reason why (\ref{eq.xrt}) is so useful is that $X^r(\omega)$ falls off as
$1/\omega$ (due to the step function in time) which makes it difficult
to obtain a faithfull representation of $X^r(t)$ from an FFT of $X^r(\omega)$. In contrast $X^{</>}(\omega)$ are well
localized (they are smooth in time), and the FFT can be safely used to
obtain $X^{</>}(\omega)$ from $X^{</>}(t)$ and vice versa. It is
possible to reduce the size of the frequency grid significantly if a
zero-padding of $X^{</>}(\omega)$ is introduced
before the FFT is applied to obtain $X^{</>}(t)$~\cite{numrecipe}.
As discussed in Sec.~\ref{sec.noneqgw}, Eq.~\ref{eq.xrt} with
$X=\Sigma$ yields the correlation part of the $GW$ self-energy. The static Hartree and
exchange terms, $\Sigma_h$ and $\Sigma_x$, are calculated from
Eqs. (\ref{eq.hr}) and (\ref{eq.exr}). Once the 
self-energies have been calculated a new set of GFs can be calculated from Eqs.~(\ref{eq.grfromdft}) and (\ref{eq.keldysh}).


\bibliographystyle{apsrev}

\begin{thebibliography}{24}
\bibitem{help1} 
We use that $(G^a)^{-1}-(G^r)^{-1}=\Sigma_{tot}^r-\Sigma_{tot}^a=\Sigma_{tot}^>-\Sigma_{tot}^<$.
\bibitem{gauss} 
These six dimensional integrals are evaluated in a two-step procedure:
First a Poisson equation for the (complex) charge distribution
$\varrho_{ik}(\bold r)=\phi_i(\bold r)^*\phi_k(\bold r)$ is solved to
obtain the potential $v_{ik}(\bold r)$. If the Poisson equation is
solved in Fourier space, one must solve for the potential $\tilde
v_{ik}$ of the neutralized charge $\tilde
\varrho_{ik}=\varrho_{ik}-\varrho_{G,ik}$, where $\varrho_{G,ik}$ is a
neutralizing Gaussian
charge distibution for which the associated potential $v_{G,ik}$ is
known analytically. In
the last step we obtain the Coulomb integral as, $V_{ij,kl}=\int
\text{d}\bold r \phi_j(\bold r)^* v_{ik}(\bold r)\phi_l(\bold r)$.
\bibitem{details}
We used the Dacapo plane wave~\cite{dacapo} code with Vanderbilt ultrasoft
pseudopotentials. The unit cell for benzene was 16\AA~in each
direction, the plane wave cut off was 340 eV, and we used the PBE xc-functional.
\end{thebibliography}

\end{document}